\documentclass[twocolumn,superscriptaddress,pra]{revtex4}

\usepackage{dcolumn}
\usepackage{bm}
\usepackage{array}
\usepackage{graphicx}
\usepackage{amsmath}
\usepackage{amssymb}
\usepackage{color}

\begin{document}


%

\title{Optomechanics of Levitated Dielectric Particles}

\author{Zhang-qi Yin}

\address{The Center for Quantum Information, Institute for Interdisciplinary Information Sciences, Tsinghua University, Beijing 100084, P. R. China\\
yinzhangqi@mail.tsinghua.edu.cn}

\author{Andrew A. Geraci}

\address{Department of Physics, University of Nevada, Reno, NV 89557\\
ageraci@unr.edu}

\author{Tongcang Li}

\address{NSF Nanoscale Science and Engineering Center, 3112 Etcheverry Hall,
University of California, Berkeley, California 94720, USA\\
tcli@berkeley.edu}


\begin{abstract}
We review recent works on optomechanics of optically trapped microspheres and nanoparticles in vacuum,
which provide an ideal system for studying macroscopic quantum mechanics and ultrasensitive force detection.
An optically trapped particle in vacuum has an ultrahigh mechanical quality factor as it is  well-isolated
from the thermal environment.  Its oscillation frequency can be tuned in real time by changing the power of
the trapping laser. Furthermore, an optically trapped particle in vacuum may rotate freely, a unique property
that does not exist in clamped mechanical oscillators.  In this review, we will introduce the current status
of optical trapping of dielectric particles in air and vacuum, Brownian motion of an optically trapped particle
at room temperature, Feedback cooling and cavity cooling of the Brownian motion. We will also discuss about using
optically trapped dielectric particles for studying macroscopic quantum mechanics and ultrasensitive force detection.
Applications range from creating macroscopic Schr{\"{o}}dinger's cat state, testing objective collapse models of
quantum wavefunctions, measuring Casimir force, searching short-range non-Newtonian gravity, to detecting gravitational waves.
\end{abstract}

\date{\today}

\maketitle

\section{Introduction}

Classical mechanics is very successful in explaining the motion of macroscopic systems, which are
are deterministic, and predictable in principle if  initial states of the system
are determined. Quantum mechanics, which explains the motion of microscopic
systems successfully, on the other hand,
is a probabilistic theory. The classical degrees of freedom, such as location and momentum,
become wavefunctions in quantum mechanics. While the  wavefunction evolves deterministically
in quantum mechanics, the wavefunctions can be in superposition states, which is the key
difference between quantum and classical physics.

Why macroscopic systems are not in quantum superpostion states? In other words, can
we observe Schr\"odinger's cat states of large objects in laboratory? This question is one of the  most outstanding
challenges in the modern physics.
We may simply explain the quantum-classical world transition by de Broglie wavelength $\lambda$,
which is defined as $\lambda=h/p$, where $h$ is Plank constant and $p$ is the momentum. Usually
the momentum of macroscopic objects is very large, and the de Broglie wavelength is too small to
be observed. Such simple explanation will predict that quantum superpositions of large objects can be observed if we can reduce the momentum $p$ to small enough values, which requires significant cooling.
Meanwhile, there are several important models proposed that the quantum-classical transition is due to more profound reasons.
For example, Penrose proposed that the conflict between general relativity and quantum mechanics leads to gravity induced collapses of quantum superpositions states\cite{Penrose96,christian2005,wezel2008}. Several other intriguing models also proposed that the collapses of massive superposition states might  intrinsically be due to  quantum mechanics being not complete\cite{diosi1989,ghirardi1986,Bassi13}. It may be necessary to introduce unknown nonlinear terms to the von Neumann equation to describe large quantum systems\cite{Nimmrichter13}. Thanks to the latest experimental developments in macroscopic quantum mechanics\cite{Chen13}, some the these models may be tested experimentally soon, which will significantly deepen our understanding of quantum mechanics.

In order to generate and observe the quantum superpostions in macroscopic systems, the momentum
(temperature) of the system should be slowed down to the quantum regime. In the past several years, the
new research area of optomechanics has had tremendous progress\cite{Chen13,Aspelmeyer2010}. Quantum ground state cooling of mechanical oscillators  by cavity cooling \cite{Wilson07,Marquardt07}
has been realized experimentally \cite{Chan11,Teufel11,Connell2010}. For the readers who are interested on the basic theory
and development of optomechanics,  please read these reviews \cite{Chen13,Aspelmeyer2010,Aspelmeyer13}.  Among the implementations of optomechanics,
optically levitated dielectric particles have attracted a lot of interest recently \cite{Chang10,Romero10,Li10,Li11,Barker10,Barker10a,Gieseler12,Kiesel13,Asenbaum13,Monteiro13}. As the objects
are levitated by optical traps, there is no mechanical contact to the environment, which is the main
decoherence source in other mechanical oscillators. Due to the absence of the mechanical contact in this system, the decoherence \cite{Yin09}
can be negligible and the oscillation frequency is fully tunable. Thus this system is ideal for study macroscopic quantum mechanics.
The center-of-mass (CoM) motion of an optically
levitated dielectric particle
could be  pre-cooled down to milli-Kelvin temperatures by feedback \cite{Li11,Gieseler12}. Then, it can be further cooled to the quantum ground state with cavity sideband cooling\cite{Chang10,Romero10,Kiesel13,Yin11,Romero11}.

After the CoM mode of an optically levitated dielectric particle is cooled down to the quantum regime,
macroscopic quantum states, such as quantum superposition states \cite{Romero11a,Romero11b}, quantum entangled
states \cite{Chang10}, and squeezed states \cite{Chang10}, may be generated.  An optically levitated dielectric particle can also be used as an ultra-sensitive detector for Casimir force,
non-Newtonian force \cite{Geraci10,PhysRevA.86.063809}, gravitational waves \cite{Arvanitaki13}, single molecules collisions \cite{Yin11} and et al..
Besides, the levitated nanoparticles are the best testbed
for gravity induced decoherence effects \cite{Penrose96,Kaltenbaek12}, which is the result of the apparent conflict between quantum mechanics
and general relativity. Beside CoM motion, the levitated dielectric particle can also rotate freely \cite{Arita11}. The rotation degree of freedom may also be used
as an resource for quantum information \cite{Law12,Shi13}.
The many body physics, such as self-assembly of the nanoparticles in vacuum, was also proposed to study \cite{Lechner12,Habraken13}.

This review is organized as follows. In Sec. 2, we will review the current status of optical trapping of dielectric particles in air and vacuum. In Sec. 3, we will discuss the CoM motion and cooling of a levitated dielectric particle.
In Sec. 4, we will discuss the macroscopic quantum mechanics of levitated dielectric particles. In Sec. 5, we will talk about the applications of a levitated dielectric particle in ultrasensitive force detection.

\section{Optical trapping of dielectric particles in air and vacuum}

\begin{figure}[b!]
\centering
\includegraphics[totalheight=5cm]{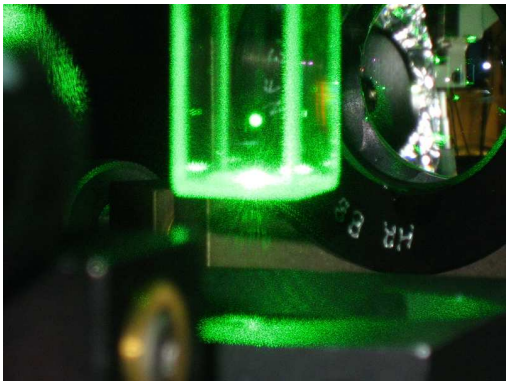}
\caption[A levitated microsphere in air]{\label{fig-III-4-1} A 4.7-$\mu$m diameter silica microsphere levitated in air inside a glass cell by an upward laser beam. The bright  dot near the center of the photo is the trapped microsphere. It appears much larger than the real size of the microsphere because of the overexposure of the camera. Figure adapted from Ref. \cite{Li13}.}
\end{figure}

Optical levitation of dielectric particles in air  by  an upward-propagating laser beam was first demonstrated by A. Ashkin and J. M. Dziedzic in 1971\cite{ashkin1971}. A few years later, optical levitation of microspheres in vacuum at pressures down to $10^{-6}$ torr was achieved \cite{Ashkin76}. An optical levitation trap is formed by the balance between the scattering force from an upward laser and the  gravitational force on a particle. A photo of a 4.7-$\mu$m diameter microsphere levitated by a laser beam in air is displayed in  Fig. \ref{fig-III-4-1}. The trapping frequency of an optical levitation trap is usually very small (about 20 Hz)\cite{Ashkin76}, which is too low for quantum ground state cooling.

\begin{figure}[tbh]
\centering
\includegraphics[totalheight=5cm]{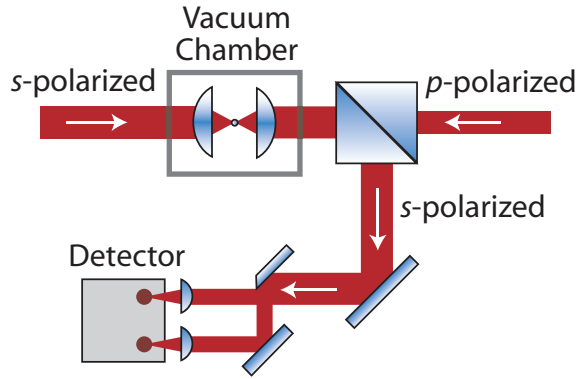}
\caption[Schematic of a dual-beam optical trap]{\label{fig-III-5} A counter-propagating dual-beam optical trap in a vacuum chamber. The Brownian motion of  a trapped particle will change the direction of the output lasers, which can be used to monitor the instantaneous position of the particle.  Figure adapted from Ref.~\cite{Li10}.}
\end{figure}

In 2010, Li \emph{et al} have trapped glass microspheres in air and high vacuum with a counter-propagating dual-beam trap\cite{Li10,Li11} (Fig. \ref{fig-III-5}). The scattering forces from the two counter-propagating beams cancel, and the gradient force forms a stable three-dimensional (3D) trap. The microspheres were initially stuck on the surface of a glass slide that was placed above the optical trap. They were launched to air by ultrasonic vibration. As they were falling down under the influence of gravity, one of them entered the optical trap and was captured. The trap was very stable and insensitive to the laser power. A 4.7 micrometer particle could be trapped stably when the power of both laser beams were changed from 5 mW to 2 W. For particles much smaller than the wavelength of the laser, the scattering force is much smaller comparing to the gradient force. Thus nano-particles may be trapped by a single tightly focused laser beam, as was demonstrated by Gieseler et al\cite{Gieseler12} recently.  A mechanical quality factor (Q) of $10^7$ has been experimentally demonstrated  at $10^{-5}$ mbar\cite{Gieseler12}, and a Q-factor of $10^8$ was recently observed at $0.5 \times 10^{-6}$ mBar\cite{Gieseler13}. These values are already higher than the quality factors achieved with clamped oscillators. In ultrahigh vacuum ($10^{-10}$ mbar) regime, the quality factor is expected to be higher than $10^{12}$.

Fig. \ref{fig-III-5} also shows a simple fast detection  system that can monitor the trajectory  of a trapped particle with ultrahigh resolution\cite{Li10}. When the trapped particle moves, it will changes the direction of the output laser slightly. Thus we can measure the particle position by monitoring the direction of one of the laser that passing through the particle.  Li et al have demonstrated a detection sensitivity of about 39~$\texttt{fm}/\sqrt{\texttt{Hz}}$ \cite{Li11}.

\section{Center-of-mass motion and cooling of a levitated dielectric particle}

\subsection{Brownian motion}
An optically trapped microsphere in  air (or a non-perfect vacuum) will exhibit Brownian motion due to collisions between the microsphere and air molecules.
The Brownian motion was discovered by Robert Brown (1773 -
1858) in 1827 when he used a simple microscope  to study the action of particles contained in the
grains of pollens.
The trajectories of a Brownian particle are commonly thought to
be continuous everywhere but not differentiable
anywhere, which means the velocity of a Brownian
particle is undefined.

In 1907, Einstein published a paper entitled ``Theoretical observations
on the Brownian motion'' in which he considered the instantaneous velocity of
a Brownian particle \cite{einstein1907}. Einstein showed that by measuring this quantity,
one could prove that ``the kinetic energy of the motion of the centre of gravity of
a particle is independent of the size and nature of the particle and independent
of the nature of its environment". This is one of the basic tenets of statistical
mechanics, known as the equipartition theorem. However, Einstein concluded
that because of the very rapid randomization of the motion, the instantaneous
velocity of a Brownian particle would be impossible to measure in practice.
In 2010, Li and et al. built a fast detection system with ultrahigh resolution and measured the instantaneous velocity of the Brownian motion  of an optically levitated  microsphere in air\cite{Li10}.

\begin{figure}[bth]
\centering
\includegraphics[totalheight=5.5cm]{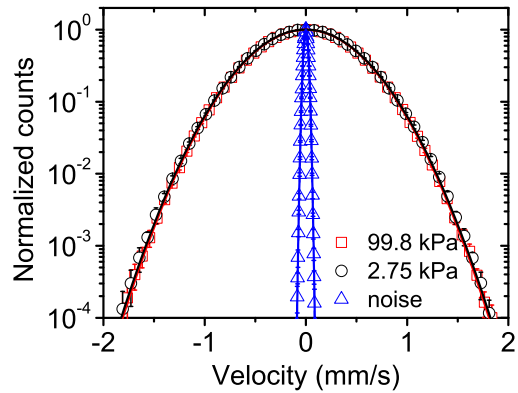}
\caption[Measured velocity distribution of a bead]{\label{measuredvelocity} The distribution of the measured instantaneous velocities of a 3 $\mu$m silica bead. The statistics at each pressure are calculated from 4 million instantaneous velocities. The solid lines are Maxwell-Boltzmann distributions.  Figure adapted from Ref.~\cite{Li10}.}
\end{figure}

The distributions of the  instantaneous velocities measured by Li et al. are displayed in Fig. \ref{measuredvelocity}. They agree  with the Maxwell-Boltzmann distribution very well. The measured rms velocities are $v_{rms}$ = 0.422 mm/s at 99.8 kPa and $v_{rms}$ = 0.425 mm/s at 2.75 kPa. These are very close to the prediction of the energy equipartition theorem, $v_{rms}=\sqrt{k_B T/M}$, which is 0.429 mm/s. As expected, the velocity distribution is independent of pressure. The rms value of the noise signal is 0.021 mm/s, which means  1.0 \r{A} spatial resolution in 5 $\mu$s. This measurement noise is about 4.8\% of the rms velocity. Fig. \ref{measuredvelocity} represents  direct verification of the Maxwell-Boltzmann distribution of velocities and the equipartition theorem of energy for  Brownian motion.

The Langevin equation of the Brownian motion of an optically trapped   microsphere is:
 \begin{equation}
\label{eq6-1}
 \frac{d^2 x_j}{dt^2} + \Gamma_0 \frac{dx_j}{dt} + \Omega^2_j x= F^{th}_j ,
\end{equation}
where $\Gamma_0$ is the viscous damping factor due to  air molecules, $\Omega_j/2\pi$ (j=1, 2, 3) are the resonant frequencies of the optical trap along the three fundamental axes (x, y, and z axes), and $F^{th}_j = \zeta_j(t) \sqrt{2 k_B T \Gamma_0 /M} $  is the Brownian stochastic force. Here $\zeta_j(t)$ is the normalized white noise process.

The damping term $\Gamma_0 \frac{dx}{dt}$ tends to stop any vibration, while the $F^{th}_j$ term drives the motion.
It is very interesting that $\Gamma_0$ is also contained in $F^{th}_j$ due to fluctuation-dissipation theorem. This keeps the average mechanical energy (kinetic and potential energy) of the microsphere to be $k_B T $ in each direction at thermal equilibrium.

 At thermal equilibrium, the power spectrum of COM motion of a trapped microsphere along each of the three fundamental mode axes is\cite{cohadon1999}:
\begin{equation}
\label{eq6-2} S_j(\omega) = \frac{2 k_B T_0}{M}  \frac{\Gamma_0}{(\Omega^2_j-\omega^2)^2+\omega^2 \Gamma^2_0} ,
\end{equation}
where  $\omega/2\pi$ is the observation frequency.

\subsection{Feedback cooling}

\begin{figure}[bth]
\centering
\includegraphics[totalheight=6cm]{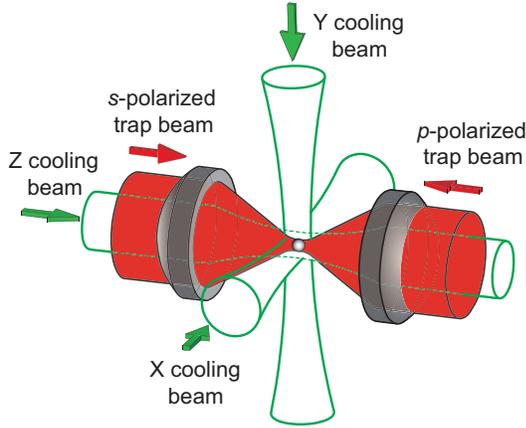}
\caption[3D feedback cooling]{\label{fig-6-feedback-loop} Simplified schematic showing a glass microsphere trapped at the focus of a counter-propagating dual-beam optical tweezer (1064 nm), and three 532nm laser beams along the axes for cooling. Figure adapted from Ref.~\cite{Li11}.}
\end{figure}

Since we can measure the instantaneous velocity of the optically trapped dielectric particle,
we can cool its CoM motion by applying a feedback force proportional to the velocity of the particle but with opposite direction (Fig. \ref{fig-6-feedback-loop}):
 \begin{equation}
\label{eq6-3}
F^{cool}_j=-\Gamma^{cool}_j \frac{dx_j}{dt}.
\end{equation}
This force will slow down the motion of the particle. With feedback cooling, the Langevin equation of the Brownian motion of an optically trapped   particle is:
 \begin{equation}
\label{eq6-4}
 \frac{d^2 x_j}{dt^2} + (\Gamma_0 + \Gamma^{cool}_j) \frac{dx_j}{dt} + \Omega^2_j x= \zeta_j(t) \sqrt{\frac{2 k_B T \Gamma_0}{M}}.
\end{equation}
In contrast to the $\Gamma_0$ due to air molecules,  $\Gamma^{cool}_j$ is only contained in the damping term but  not in the heating term.
Let $\Gamma^{tot}_j=\Gamma_0+\Gamma^{cool}_j$ be the total damping factor, and $T^{cool}_j=T_0 \Gamma_0/\Gamma^{tot}_j$ be the effective temperature of the motion with feedback cooling,  the power spectrum with feedback cooling can be rewritten as:
\begin{equation}
\label{eq6-6} S^{cool}_j(\omega) = \frac{2 k_B T^{cool}_j}{M}  \frac{\Gamma^{tot}_j}{(\Omega^2_j-\omega^2)^2+\omega^2 (\Gamma^{tot}_j)^2},
\end{equation}
which has the same form as Eq. \ref{eq6-2}. Because the effective temperature is $T^{cool}_j=T_0 \Gamma_0/(\Gamma_0+\Gamma^{cool}_j)$ , the motion  can be cooled significantly by applying a feedback damping $\Gamma^{cool}_j>>\Gamma_0$.

\begin{figure}[bth]
\centering
\includegraphics[totalheight=6cm]{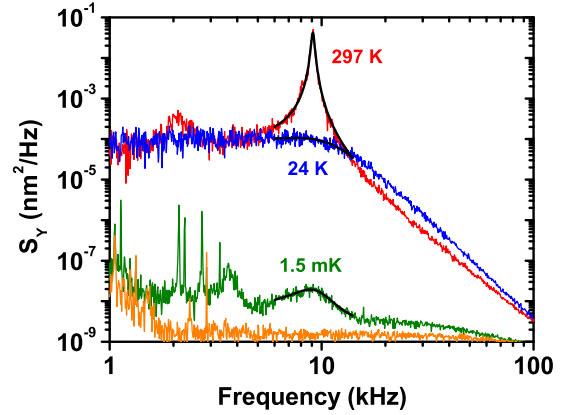}
\caption[Power spectra of a bead along Y axis]{\label{fig-VI-3D-feedback-result-y}Power spectra of a trapped 3-$\mu$m diameter microsphere along the  Y  axis as it is cooled. The red curve is the intrinsic spectrum at 637 Pa without feedback cooling, the blue curve is the spectrum at 637 Pa with feedback cooling,
the green curve is the spectrum at 5.2 mPa with feedback cooling, and the orange curve is  the noise signal when there is no particle in the optical trap. The black curve is the fit of a thermal model (see text for details). The mode temperatures are obtained from these fits. Figure adapted from Ref.~\cite{Li11}.}
\end{figure}

Figure \ref{fig-VI-3D-feedback-result-y}  show experimental results of feedback cooling along Y axis by Li et al \cite{Li11}. Before feedback is turned on, the resonant frequencies ($\omega_j/2 \pi$) are $ 8066 \pm 5$ Hz, $ 9095 \pm 4$ Hz, and $ 2072 \pm 6$ Hz for the  fundamental modes  at 637 Pa along the X, Y, and Z axes, respectively.  After the feedback cooling circuits were turned on, the temperature of the Y mode changed from 297 K to 24 K at 637 Pa. The mode temperature is obtained by fitting the measured power spectrum with Eq. \ref{eq6-6}.
Then Li et al reduced the air pressure while keeping the feedback gain almost constant,
 thus the heating rate  due to collisions from air molecules  decreases,  while the cooling rate remains constant. As a result, the temperature of the motion dropped. At 5.2 mPa, the mode temperatures were $150 \pm 8$ mK, $1.5 \pm 0.2$ mK, and $68 \pm 5$ mK for the x, y and z modes.
The mean thermal occupation number $\langle n \rangle = k_B T^{fb}_j/\hbar \omega_j$ of the y mode is reduced from about $6.8\times10^8$ at 297K to about $3400$ at 1.5 mK.

\subsection{Cavity sideband cooling}\label{sec:cooling}

 In 2009, two groups
proposed to use the cavity sideband cooling scheme \cite{Wilson07,Marquardt07} to
cool the CoM mode of optically levitated nanoparticle down to the ground state \cite{Chang10,Romero10}. Recently,
the cavity cooling was partially realized by Kiesel et al.\cite{Kiesel13} and Asenbaum et al.\cite{Asenbaum13}.
Kiesel et al. have optically trapped a nanoparticle inside the optical cavity, and achieved the sideband limit\cite{Kiesel13}.
Because of the relatively high pressure (4 mbar) in their experiment, they were only able to cool the effective temperature of a levitated nanoparticle from room temperature to about 64~K.
They believe that the quantum
ground state may approach, if they can increase the vacuum to $10^{-7}$ mbar. Here, we will
give a short review on the theory of cavity sideband cooling of optically trapped nanoparticle.

\begin{figure}
  \centering
  \includegraphics[width=7cm]{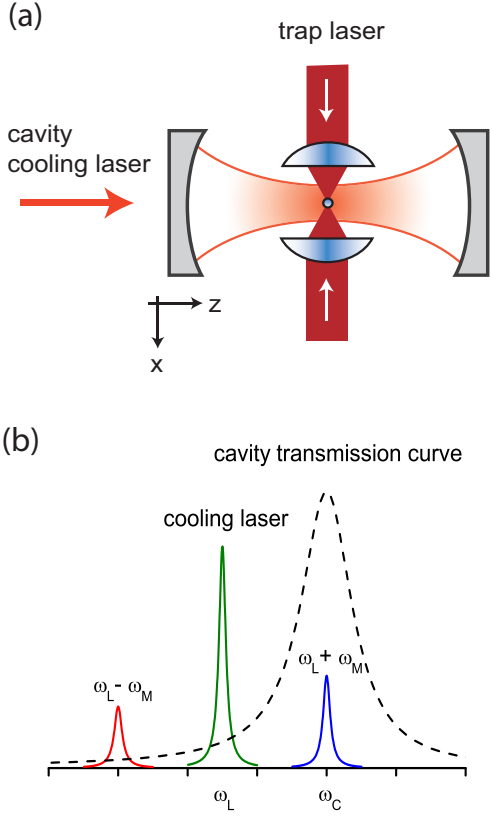}\\
  \caption{(a) Scheme of 1D cavity cooling. A nanoparticle is trapped inside
an optical cavity with a dual-beam trap. (b) Principle of 1D cavity cooling.
The frequency of the cooling laser $\omega_l$ is slightly smaller than the resonant
frequency of the optical cavity $\omega_c$. The mechanical vibration of the trapped
microsphere at frequency $\omega_m$ induces two side bands of the laser at frequencies
of $\omega_l+\omega_m$ and $\omega_l-\omega_m$.}\label{fig:cooling}
\end{figure}

The typical scheme of cavity sideband cooling for optically trapped nanoparticle is show in Fig. \ref{fig:cooling}.
The optical trap is placed in the optical cavity, and a nanosphere with mass $m$ is loaded
in the trap.
We denote vibration frequency of the nanoparticle along the $z$ axis be $\omega_m$,
the frequency of the cooling laser as $\omega_l$, the resonant frequency of the cavity as $\omega_c$, the
intrinsic cavity linewidth as $\kappa$ and the rate of a photon scattered by the microsphere as $\gamma_{sc}$.
The sideband cooling requires that
the linewidth $\kappa$ is much shorter than the optical trap frequency $\omega_m$. As the size of nanoparticle
is much less than the wavelength of the trapping and cooling light, and the molecule collision rate is very low at high vaccum,
the photon scattering decoherence rate
$\gamma_{rc}$ is usually much less than $\kappa$, and can be neglected.

Let us consider a nanoparticle at position $z$ moving with momentum $p$ along the $z$ axis inside of a driven
cavity. The nanoparticle  causes the cavity frequency to shift by an amount
$$ \delta \omega_c = -\frac{1}{2} \frac{\int d^3 {\bf r}
    \delta P({\bf r}) \dot {\bf E}({\bf r}) }{\int d^3{\bf r} \epsilon_0
    {\bf E}^2({\bf r})} \cdot \omega_{c0},
$$
where $\omega_{c0}$ is the resonant frequency of a cavity without the nanosphere, ${\bf E}({\bf r})$ is the
cavity mode profile and $\delta P({\bf r})$ is the variation in permittivity induced by the nanosphere. Due to
the tiny scale of the nanosphere (much less than laser wavelength), we can use can use Rayleigh approximation, and
have $P({\bf r}')\simeq \alpha_{\mathrm{ind}}
E({\bf r})\delta ( {\bf r} - {\bf r}')$, with ${\bf r}$ the CoM position of the nanosphere,
$\alpha_{\mathrm{ind}} = 3 \epsilon_0 V(\frac{\epsilon-1}{\epsilon+2})$ the polarizability, $V$ the
sphere volume, and $\epsilon$ is the electric permittivity.

The Hamiltonian of the system can be approximated as
\begin{equation}\label{eq:Heff}
  H_{eff} =    \hbar \omega_m
a_m^\dagger a_m -\hbar \Delta_c a^\dagger_{c} a_{c} +
\frac{\hbar\Omega_c}{2} ( a_{c} +a_{c}^\dagger) + \hbar g_j
a^\dagger_{c} a_{c} (a_m+ a_m^\dagger),
\end{equation}
where $g_j=q_{\mathrm{zpf}j} \partial U(z)/\partial j|_{z=z_0}$ characterizes the coupling strength between
the cavity  mode and the oscillation of the nanosphere, $U(z)$ is the nanoparticle induced
frequency shift, and $z_{\mathrm{zpf}}= \sqrt{\hbar/2m\omega_j}$ is zero-point
fluctuation for the phonon mode $a_m$. $\Delta_c=\omega_c-\omega_l$ is detunings between the
lasers and the cavity modes $a_c$. $\Omega$ is the driving strength of the cooling laser.

From Eq. \eqref{eq:Heff}, the linearized Langevin equations of motion for
our system are,
\begin{equation}
  \label{eq:Langevin}
 \begin{aligned}
  \dot{a_{c}}=& (i\Delta_{c}' - \kappa_j/2)a_{c} - ig \alpha_c
                  (a_m + a_m^\dagger) +\sqrt{\kappa_j} a_{c}^{\mathrm{in}},\\
 \dot{a_m} = &-i\omega_m  a_m - i g (\alpha_c a_{c}^{\dagger} + \alpha_m^* a_{c}),
 \end{aligned}
\end{equation}
where $\alpha_c = i\Omega/(2i\Delta_{c}' -\kappa)$,
$\Delta_{c}' = \Delta_{c} + 2g_j^2|\alpha_c|^2/\omega_{m}$,
$\alpha_c$ is the amplitude of cavity mode $a_{c}$, and $\Delta'_{c}$
is the effective detuning between the driving laser and the cavity
mode $a_{c}$.
The linearization of the Langevin equations is valid only
if the state is stable. The stable criteria is \cite{Yin11}
$  S_1 = 4\Delta_{c}' \omega_m g^2 \alpha_c^2 \kappa^2>0,
 S_2 = \omega_m \Delta_{c}{'^2}-g^2\alpha_c^2 \Delta_{c}'>0.
$
Because of $\Delta_{c}'>0$, the criteria $S_1$ is always valid. The criteria $S_2$
are valid only when $g \alpha_c <\sqrt{\omega_m \Delta_{c}'}$.
 To realize resolved sideband cooling, we require
$\omega_m \gg \kappa$. We suppose $ |g \alpha_c| \ll \kappa$, and
find that the final phonon number is 
$$
n_{m}= -\frac{(\omega_m+\Delta'_{c})^2 +(\kappa/2)^2}{4\omega_m
\Delta'_{c}}.
$$
In the special case of $\Delta_{c}'= -\omega_m$, the final phonon
number is $n_{m} =(\kappa/4\omega_m)^2 \ll 1$. The cooling rate is
$\Gamma =  g^2 |\alpha_c|^2 /[\kappa(1+\frac{\kappa^2}{16\omega_m^2})]$.

\subsection{3D sideband cooling}

A nanoparticle will scatter the trapping/cooling laser to all three dimensions and
cause 3D heating. 
In order to
achieve ground state cooling of an optically trapped nanosphere, we must use
a 3D cooling scheme. We can add two more cavities for cooling the other two
dimensions, but the system will become too complex to be realized experimentally.
A better method to cool and measure the 3D motion of a nanosphere
is to use the TEM00, TEM01, and TEM10 modes of a single cavity, as proposed by Yin \emph{et al}. \cite{Yin11}.
The TEM01 and TEM10 beams can be generated from a TEM00 beam by two
phase plates. Each one of these three modes can be coupled to the motion
of a trapped nanosphere along one orthogonal axis.

\begin{figure}[htbp]
\centering
\includegraphics[width=8cm]{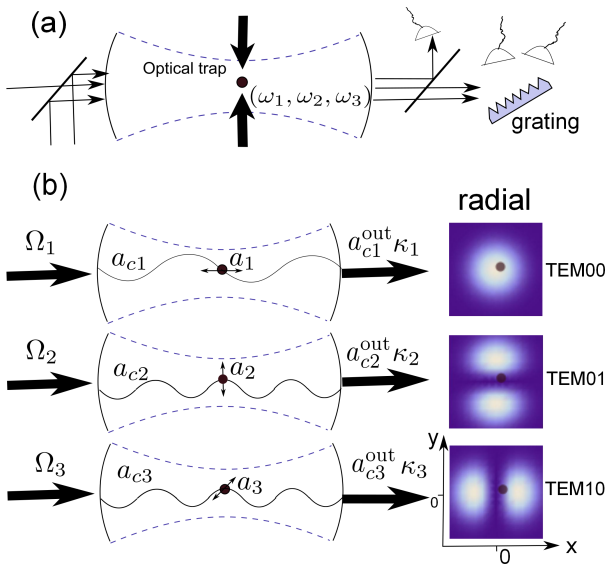}
\caption[3D sideband cooling]{\label{fig:3Dcooling} (a) Cooling and detecting scheme. A
nanosphere is trapped by a dual-beam optical tweezer inside of a
cavity. The cavity is driven by three lasers in TEM00, TEM01 and
TEM10 modes. The TEM01 mode laser has different polarization, and is
separated from the other two lasers by a polarizing beam splitter
for detection. The TEM00 and TEM01 lasers have different
frequencies, and are separated by a grating for detection. (b) Three
cooling modes TEM00, TEM01, and TEM10, and their radial
distribution. The black dot represents the position of a trapped
nanosphere. Figure adapted from Ref. \cite{Yin11}}
\end{figure}

As shown in Fig. \ref{fig:3Dcooling}, we consider an optically trapped nanosphere
with mass $m$ confined in a cavity by means of an optical tweezer
\cite{Li10}.
The frequencies of the optical trap along the $z$, $x$, and $y$ axes are
$(\omega_1$, $\omega_2$, and $\omega_3)$. Beside the
conventional method of using a cooling laser with TEM00 mode to cool
the motion along $z$ direction, we add two non-Gaussian beams with
TEM01 and TEM10 modes to drive the cavity in order to cool the
motion along the $x$ and $y$ directions, respectively. The resonant
frequencies of the cavity modes $a_{c1}$, $a_{c2}$, and $a_{c3}$ are
$\omega_{c1}$, $\omega_{c2}$, and $\omega_{c3}$, respectively. The
detunings between the lasers and the cavity modes are $\Delta_{cj} =
\omega_c^j - \omega_L^j$ $(j=1,2,3)$. We suppose that the TEM01 and
TEM10 lasers have the same frequency, but with orthogonal
polarization. The TEM00 and TEM01 lasers have the same polarization,
but different frequencies. In practical, the frequency differences between TEM00 and
TEM01 (TEM10) could be very large, and the TEM01 and TEM10 modes are orthogonal in
polarizations. Therefore the interference between the three cavity modes
can be neglected.

The total Hamiltonian of the system in the rotating frame is\cite{Yin11}
\begin{equation}\label{Hamiltonian}
  \begin{aligned}
    H = &\sum_{j=1}^3 \big[\hbar \omega_j a_j^\dagger a_j -
     \hbar (\Delta_j-U_j) a^\dagger_{cj} a_{cj} +
   \frac{\hbar\Omega_j}{2} ( a_{cj} +a_{cj}^\dagger)\big],
  \end{aligned}
\end{equation}
where $a_j$ characterizes the phonon mode along $q_j$ direction with
$q_1=z, q_2=x, q_3=y$. $\Omega_j$ is the driving strength by the
lasers and $U_j$ characterizes the coupling between the cavity mode
$a_{cj}$ and the nanosphere.
 In the limit that
$\epsilon \gg 1$, where $\epsilon$ is the electric permittivity of
the nanosphere, we get \cite{Chang10}
\begin{equation*}
   \begin{aligned}
    U_1=& -  \frac{3V}{2V_{c1}}
\exp (-\frac{2x^2+ 2y^2}{w^2}) \cos^2 (k_1z +\varphi_1)
\omega_{c1},\\
    U_2 = &-  \frac{3V}{2V_{c2}}
\frac{x^2}{w^2}\exp (-\frac{2x^2+ 2y^2}{w^2}) \cos^2 (k_2z +
\varphi_2)\omega_{c2},\\
    U_3 = &- \frac{3V}{2V_{c3}} 
\frac{y^2}{w^2}\exp (-\frac{2x^2+ 2y^2}{w^2})\cos^2 (k_3z +
\varphi_3)\omega_{c3},
   \end{aligned}
\end{equation*}
with $V_{c1}= (\pi/4)Lw^2$ and $V_{c2}=V_{c3}=(\pi/16)Lw^2$.

We assume the optical tweezer to be much stronger than the
cavity-mode-induced trap, and neglect the effects of cooling lights
on trapping. Besides, if we carefully choose the location of the
trap, such as $z_0=0$, $x_0=y_0=0.25w$, $\varphi_1=\pi/4$, and
$\varphi_2=\varphi_3=0$, the gradients of the three light fields lie
approximately along the three axes. The effective Hamiltonian is
\begin{equation}
 \begin{aligned}
 \label{eq:effH1}
 H_{eff} = &  \sum_{j=1}^{3} \big[\hbar \omega_j
a_j^\dagger a_j -\hbar \Delta_j a^\dagger_{cj} a_{cj} +
\frac{\hbar\Omega_j}{2} ( a_{cj} +a_{cj}^\dagger) \\
          & + \hbar g_j
a^\dagger_{cj} a_{cj} (a_j+ a_j^\dagger) \big],
 \end{aligned}
\end{equation}
where $g_j=q_{\mathrm{zpf}j} \partial U(x,y,z)/\partial
j|_{x=x_0,y=y_0,z=z_0}$ characterizes the coupling strength between
the cavity  mode and the oscillation of the nanosphere, and
$q_{\mathrm{zpf}j}= \sqrt{\hbar/2m\omega_j}$ is zero-point
fluctuation for the phonon mode $a_j$.  In general, $g_1$ can be one
to two orders larger than $g_2$ and $g_3$.

As the 3D motional modes of the system are decoupled with each other in effective
Hamiltonian \eqref{eq:effH1}, we can find the final phonon number equation with the similar method
discussed in the previous subsection \ref{sec:cooling}.
$
n_{mj}= -\frac{(\omega_j+\Delta'_{cj})^2 +(\kappa_j/2)^2}{4\omega_j
\Delta'_{cj}}.
$
In the special case of $\Delta_{cj}= -\omega'_j$, the final phonon
number is $n_{mj} =(\kappa_j/4\omega_j)^2 \ll 1$.

\subsection{Noise and decoherence}

Here we briefly discuss the noise and decoherence in optically levitated nanoparticles system.
The dominant noise sources for the CoM mode of nanoparticles are collisions
with a background gas and momentum recoil kicks due to scattered
photons. The noise contributions from shot noise, blackbody radiation are
negligible \cite{Chang10}. For collisions with a background gas,
it is found that the the damping rate of the phonon is $\gamma_g =(16/\pi) (P/vr\rho)$,
where $P$ and $v$ are background gass pressure and mean speed, $r$ is the
radius of the sphere, and $\rho$ is the density of the nanosphere.
For $\omega_m=0.5$ MHz, $r=50$ nm, room temperature gas with $P=10^{-10}$ Torr,
we find that $\gamma_g=10^{-6} \text{s}^{-1}$. Therefore, the molecules collision induced
decoherence is also very small. In fact, we can directly measure this collision
by output mode \cite{Yin11}. We will discuss this in the next section.
Photon scattering will entangle the mechanical mode and output light, and leads to
heating of the mechanical mode, too. Considering motion only along the z direction,
it is found that \cite{Chang10} $\gamma_{sc} = (2/5) (\omega_r/\omega_m) R_{sc}$,
where $\omega_r=\hbar k^2 /2\rho V$ is the recoil frequency, $R_{sc} = 24\pi^3 \frac{I_0}{\lambda^4}
\frac{V^2}{\hbar \omega_c} (\frac{\epsilon-1}{\epsilon +2})^2$ is the photon scattering
rate for sphere. The photon scattering rate could be very large
( $R_{sc} \sim 10^{14} \text{s}^{-1}$ for $I_0 = 1 \text{W}/ \mu\text{m}^2$ and $r=50$ nm), while the momentum of photons is much smaller than that of background air molecules. It is convenient to define a dimensionless parameter
$\phi = \frac{\gamma_{sc}}{\omega_m}= \frac{4\pi^2}{5} \frac{\epsilon-1}{\epsilon+2} (V/ \lambda^3)$ \cite{Chang10},
which can be much less than $1$ if $V$ of nanophere is much
less than $\lambda^3$. Therefore, if we want to decrease the photon recoil heating, we should
trap smaller nanoparticle, or use trapping and cooling laser with longer wavelength.
We may also use magnetic force to trap the nanoparticle, where the photon scattering effect is negligible \cite{Isart12,Cirio12}.

Then we consider the
heating effects from the optical trap \cite{PhysRevA.56.R1095}. 
The heating mainly comes from the laser intensity
fluctuation and the laser-beam-pointing noise. For the former, we
define the fluctuations of the laser $\epsilon(t) = (I(t)-I_0)/I_0$,
with $I_0$ the average intensity and $I(t)$ the laser intensity at
time $t$. By using first-order time-dependent perturbation theory,
we get  $\langle \dot{E}\rangle = \frac{\pi}{2} \omega_j^2
S_\epsilon (2 \omega_j) \langle E\rangle$ \cite{PhysRevA.56.R1095}.
The heating constant is $\Gamma_\epsilon = \frac{\pi}{2} \omega_j^2
S_\epsilon (2\omega_j)$, where $S_\epsilon (\omega) = \frac{2}{\pi}
\int^\infty_0 d\tau \cos (\omega\tau) \langle \epsilon(t)
\epsilon(t+\tau)\rangle$ is the one-sided power spectrum of the
fractional intensity noise, which could be on the order of
$10^{-14}\mathrm{Hz}^{-1}$. For the trap frequency of MHz,
$\Gamma_\epsilon$ approaches the order of $10^{-1}$Hz. The
laser-beam-pointing noise is originated from the fluctuation
relevant to the location of the trap center, which is independent of
the phonon energy. Similarly, we may get $\langle\dot{E} \rangle =
\frac{\pi}{2} m \omega_j^4 S_j (\omega_j),$ where $j=x,y,z$, and
$S_j(\omega)$ is the noise spectrum of location fluctuations. We
define the heating rate as $ \Gamma_j = \frac{\pi}{2} m \omega_j^4
S_j (\omega_j) /(\hbar \omega_j)$, which represents phonon number
increase per second. If we set $\Gamma_j$ to be on the order of
$10^{-1}$Hz, we should make sure that $S_j (\omega_j)$ is around
$10^{-35}\mathrm{m}^2/$Hz for $\omega_j\sim 1$MHz. Experimentally
$S_j(\omega)$ has been controlled less than $10^{-34} \mathrm{m}^2/$
Hz for $\omega \sim 2\pi$ kHz \cite{PhysRevLett.99.160801}. With the
increase of the optical trap frequency to large detuning from the
system's resonant frequency, $S_j(\omega_j)$ is dropping down
quickly. Therefore, we believe that the laser-beam-pointing noise
could be well controlled and the heating rate $\Gamma_j$ would be
less than $0.1$ Hz.

The phase noise induced by the cooling laser also need to be
seriously considered
\cite{PhysRevA.78.021801,PhysRevA.80.063819,Yin09}. Because the
cooling laser is of finite linewidth, the laser field can be wrote
down as $\varepsilon (t) = \varepsilon e^{i\phi(t)}$. We assume the
phase noise $\phi(t)$ to be Gaussian and with zero mean value. For
the Lorentzian noise spectrum with $S_{\dot{\phi}}(\omega) =
2\Gamma_L \gamma_c/(\gamma_c^2 + \omega^2)$, and correlation
function $\{ \dot{\phi(s)} \dot{\phi(s')} \} = \Gamma_L \gamma_c
\exp (-\gamma_c| s-s'|)$, where $\Gamma_L$ is the linewidth of the
laser and $\gamma_c^{-1}$ is the correlation time of the laser phase
noise, the phonon number limited by this noise is
 $n_{ph} > n_c \frac{\Gamma_L}{\kappa}
\frac{\gamma_c^2}{\gamma_c^2+ \omega_j^2}$
\cite{PhysRevA.80.063819}. If we choose $\Gamma_L =1$ kHz, $\gamma_c
= 3$ kHz, $\omega_j=10^6$ Hz, and $n_c=10^7$,  we have $n_{ph} \ll
1$. Besides, we may use double resonance
scheme to further increase the cooling rate and suppress the phase noise
\cite{Yin09,Pender12}.

\section{Macroscopic quantum mechanics}

After the CoM mode of optically trapped nano(micro)-particle being cooled to the quantum ground state,
a lot of quantum states can be generated, and many interesting quantum phenomena could be
observed in this macroscopic systems. In this section, we will summarize the recent developments in
this direction.

\subsection{State transfer and applications}
In the previous section, we focused on the CoM mode of the nanoparticle and calculated the
steady phonon number when the cooling laser is on. In fact, the cooling laser also realizes the
quantum interface between cavity mode and the phonon mode. With the interface, we can archive quantum
state transfer between the cavity and phonon modes \cite{Chang10}.  From Eq.\eqref{eq:Langevin},
we have a reduced equation
under rotating wave approximation, in the case of $\Delta_{cj}' =
-\omega_m$ and $\omega_m \gg \kappa, \alpha_c g$, as \cite{Yin11},
\begin{equation}
  \label{eq:Langevin1}
   \begin{aligned}
  \dot{a_{c}}=&  - \frac{\kappa}{2}a_{c} - ig \alpha_c
                  a_m  +\sqrt{\kappa} a_{c}^{\mathrm{in}},\\
 \dot{a_m} = &- i g \alpha_c a_{c}.
 \end{aligned}
\end{equation}
In the limit $\kappa \gg g\alpha_c$, using boundary condition
$a_{c}^{\mathrm{out}} = -a_{c}^{\mathrm{in}} + \sqrt{\kappa}
a_{c}$, we get $a_{c}^{\mathrm{out}} = -i
\frac{2g\alpha_c}{\sqrt{\kappa}} a_m + a_{c}^{\mathrm{in}}$,
$\dot{a}_m = -\frac{2g^2 \alpha_c^2}{\kappa} a_m
-\frac{2ig\alpha_c }{\sqrt{\kappa}} a_{c}^{\mathrm{in}}$.
Therefore the CoM motion of the nanosphere can be  mapped to the cavity output fields.
Physically, the cooling mechanism can be viewed as transferring phonon excitation
to the cavity mode, and finally leaking out of cavity.

The quantum state transfer between photon and phonon modes has many applications. The first one is generating
superposition state $(|0\rangle +|1\rangle)/\sqrt{2}$ \cite{Romero10}, where $|0\rangle$ ($|1\rangle$)
is the ground state (first Fock state) of the CoM phonon mode. We impinge a single-photon state
into the cavity. Part of the photon will reflect, and part of it transmit. In present of cooling laser,
the Langevin equations \eqref{eq:Langevin1} swap the photon state into the CoM phonon state of nanoparticle.
We the entangle state $|0\rangle_r|1\rangle_m + |1\rangle_r |0\rangle_m$, where $r$ denotes the reflecting
photon mode, and $m$ denotes the CoM phonon mode. The motional state collapses into the superposition state
$\Psi \rangle = c_0 |0\rangle_m + c_1 |1\rangle_1$, by performing a balanced homodyne measurement and by switching off the driving field.
Here the coefficients $c_{0(1)}$ depend on the measurement result. This state can be detected by transferring it back to
the cavity with a red-detuning laser and then performing tomography on the output field.

The second application of the quantum state transfer is generating the squeezed state of light \cite{Chang10}.
In order to generate the squeezed light we need to creat the mechanical squeezed state, then transfer the
squeezed properties to the output light by quantum state transfer. We add a sinusoidally varying component
to the intensity of the trapping beam, which yields the Hamiltonian
of a parametric amplifier
\begin{equation}\label{eq:squeezed}
  H_s= \epsilon_m \omega_m^2 z^2 \sin 2\omega_m t.
\end{equation}
Here $\epsilon_m$ is a small parameter characterizing the strength of the modulation
of the trap frequency. We are interested in the outgoing light over a
narrow frequency range near the cavity resonance, specifically
considering $\mathbf{X}_{\pm,\mathrm{out}}(\omega=0)$. Taking the limit as one
approaches threshold and $\Gamma=\kappa$, the variance in the output light
is given by \cite{Chang10}
$$
\Delta \mathbf{X}^2_{+,\mathrm{out}} (\omega=0) = \frac{5}{16} \frac{\kappa^2}{\omega_m^2}.
$$
Here we neglect the phonon recoil heating effects. We find that the output squeezing could
easily approach to $30$ dB of noise reduction relative to the vacuum state.

We can use quantum state transfer to realize entanglement transfer between remote
nanoparticles trapped in separate cavities \cite{Chang10}, or to generate the
Schr\"odinger¡¯s cat state $|\alpha\rangle + |- \alpha\rangle$ \cite{Romero10}.
Besides, if we drive the light on resonance of blue sideband, we will generate
two-mode squeezed state between phonon and photon modes. By combing blue and red sideband
driving, we can generate two-mode squeezed light with the method similar as the Ref. \cite{YH09}.
However, as there is no non-linear coupling in CoM mode of trapped nano-particles (at least for the first order),
the non-Gaussian state of mechanical mode can only be generated by mapping the
photon state into it. Therefore, the quantum state that can be generated(detected) in trapped
nanoparticle is depending on the input(output) state of light, which highly limits the applications
of the system.

\subsection{Optically trapped nanoparticle with built-in spins}

\begin{figure}[tbph]
\centering
\includegraphics[width=7cm]{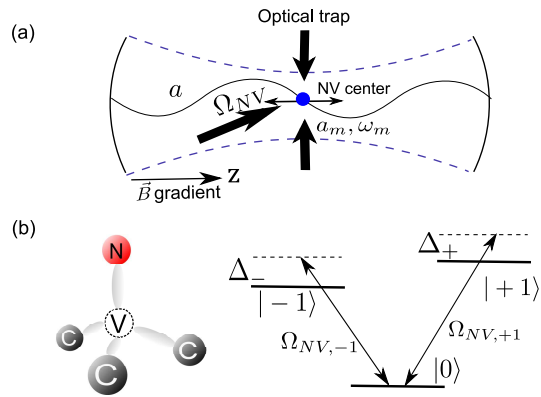}
\caption[Levitated nanodiamond with building in NV center spin]{\label{fig:nanodiamond} (a) A nanodiamond with a NV center is optically
trapped in vacuum with spin-mechanical coupling enabled through a nearby
magnetic tip and opto-mechanical coupling through a cavity around. (b) The
atomic structure (left) and the level diagram (right) in the ground state
manifold for a NV center in the nanodiamond. Figure adapted from Ref. \cite{Yin13}.}
\end{figure}

 Nanodiamonds with NV centers have been
recently trapped by optical tweezers in fluid \cite{Horowitz12,Geiselmann13} and
atmospheric air \cite{Neukirch13}, and similar technologies can be used to optically
trap them in vacuum \cite{Gieseler12}. The nonlinear interaction required for generation
of non-Gaussian quantum states is enabled through the spin-mechanical coupling with a
built-in nitrogen-vacancy center inside the nanodiamond \cite{Yin13}. By detecting
the spin state of NV center in nanocrystal diamond, the phonon state can be detected without cavity mode.

As shown
in Fig. \ref{fig:nanodiamond}, we consider a nanodiamond of mass $m$ optically trapped in vacuum
with trapping frequency $\omega _{m}$. The motion of its CoM mode
$a_{m}$ is described by the Hamiltonian $H_{m}=\hbar \omega
_{m}a_{m}^{\dagger }a_{m}$. The nanodiamond has a built-in NV center with
its level configuration shown in Fig.1b in the ground-state manifold. The NV
spin is described by the Hamiltonian $H_{\mathrm{NV}}=\hbar (\omega
_{+1}|+1\rangle \langle +1|+\omega _{-1}|-1\rangle \langle -1|)$ , where we
have set $|0\rangle $ as the energy zero point. A magnet tip near the NV
center induces a strong magnetic field gradient \cite{Mamin07}, which
couples the electron spin and the CoM oscillation of the
nanodiamond. The coupling Hamiltonian is denoted by $H_{\mathrm{NV}m}=\hbar
\lambda S_{z}(a_{m}+a_{m}^{\dagger })$ \cite{Rabl09,Kolk12}, where $%
S_{z}\equiv |+1\rangle \langle +1|-|-1\rangle \langle -1|$. The coupling
strength $\lambda =g_{s}\mu _{B}G_{m}a_{0}/\hbar $, where $a_{0}=\sqrt{\hbar
/2m\omega _{m}}$, $g_{s}\simeq 2$ is the Land\'{e} g-factor, $\mu _{B}$ is
the Bohr magneton, and $G_{m}$ is the magnetic field gradient along the NV
center axis.


In order to
prepare the Fock states, we first cool the mechanical mode to the ground
state with sideband cooling \cite{Chang10}, or spin assistant phonon cooling \cite{Rabl09}.
The NV spin is initially set to the state $|0\rangle $, which is
decoupled from the mechanical mode during the cooling. Initialization and
single shot detection of the NV spin have been well accomplished
experimentally \cite{Robledo11}. We assume that the NV\ center is at a
position with zero magnetic field and a large field gradient. We apply a
microwave drive with the Hamiltonian $H_{drive}=\hbar (\Omega _{\mathrm{NV}%
,+1}e^{i\omega _{l+}t}|0\rangle \langle +1|+\Omega _{\mathrm{NV}%
,-1}e^{i\omega _{l-}t}|0\rangle \langle -1|+h.c.)/2$ and set the Rabi
frequency $\Omega _{\mathrm{NV},\pm 1}=\Omega _{\mathrm{NV}}$ and the
detuning $\Delta _{\pm }\equiv \omega _{l\pm }-\omega _{\pm 1}=\Delta $.
With $\Delta \gg |\Omega _{\mathrm{NV}}|$, we adiabatically eliminate the
level $|0\rangle $ and get the following effective Hamiltonian
\begin{equation}
\begin{aligned} H_{e} =\hbar \omega_m a_m^\dagger a_m +\hbar\Omega\sigma_z +
\hbar \lambda (\sigma_+ + \sigma_-) (a_m +a_m^\dagger), \end{aligned}
\label{eq:Heff1}
\end{equation}%
where $\Omega =|\Omega _{\mathrm{NV}}|^{2}/4\Delta $, $\sigma _{z}=|+\rangle
\langle +|-|-\rangle \langle -|$, $\sigma _{+}=|+\rangle \langle -|$, $%
\sigma _{-}=|-\rangle \langle +|$, and we have defined the new basis states $%
|+\rangle =(|+1\rangle +|-1\rangle )/\sqrt{2}$, $|-\rangle =(|-1\rangle
-|-1\rangle )/\sqrt{2}$. In the limit $\lambda \ll \omega _{m}$, we set $%
\Omega =\omega _{m}/2$ and use the rotating wave approximation to get an
effective interaction Hamiltonian between the mechanical mode and the NV
center spin, with the form $H_{JC}=\hbar \lambda \sigma _{+}a_{m}+h.c.$.
This represents the standard Jaynes-Cummings(J-C) coupling Hamiltonian.
Similarly, if we set $\Omega =-\omega _{m}/2$, the anti J-C Hamiltonian can
be realized with $H_{aJC}=\hbar \lambda \sigma _{+}a_{m}^{\dagger }+h.c.$.

Arbitrary Fock states and their superpositions can be prepared with a
combination of J-C and anti J-C coupling Hamiltonians. For example, to
generate the Fock state $|2\rangle _{m}$, we initialize the state to $%
|+\rangle |0\rangle _{m}$, turn on the J-C coupling for a duration $%
t_{1}=\pi /(2\lambda )$ to get $|-\rangle |1\rangle _{m}$, and then turn on
the anti J-C coupling for a duration $t_{2}=t_{1}/\sqrt{2}$ to get $%
|+\rangle |2\rangle _{m}$. The Fock state with arbitrary phonon number $%
n_{m} $ can be generated by repeating the above two basic steps, and the
interaction time is $t_{i}=t_{1}/\sqrt{i}$ for the $i$th step \cite%
{Meekhof96}. Superpositions of different Fock states can also be generated.
For instance, if we initialize the state to $(c_{0}|+\rangle +c_{1}|-\rangle
)\otimes |0\rangle _{m}/\sqrt{2}$ through a microwave with arbitrary
coefficients $c_{0},c_{1}$, and turn on the J-C coupling for a duration $%
t_{1}$, we get the superposition state $|-\rangle \otimes (c_{1}|0\rangle
_{m}+ic_{0}|1\rangle _{m})/\sqrt{2}$.
Using the optical cavity, the Fock
state $|n_{m}\rangle _{m}$ of mechanical mode can also be mapped to the
corresponding Fock state of the output light field \cite{Yin11}.

The effective Hamiltonian for the spin-phonon coupling takes the form $%
H_{QND}=\hbar \chi \sigma _{z}a_{m}^{\dagger }a_{m}$ with $\chi =4\Omega
\lambda ^{2}/(4\Omega ^{2}-\omega _{m}^{2})$ when the detuning $||\Omega
|-\omega _{m}/2|\gg \lambda $. The Hamiltonian $H_{QND}$ can be used for a
quantum non-demolition measurement(QND) measurement of the phonon number: we
prepare the NV center spin in a superposition state $|+\rangle +e^{i\phi
}|-\rangle )/\sqrt{2}$, and the phase $\phi $ evolves by $\phi (t)=\phi
_{0}+2\chi n_{m}t$, where $n_{m}=a_{m}^{\dagger }a_{m}$ denotes the phonon
number. Through a measurement of the phase change, one can detect the phonon
number.

The preparation and detection of the Fock states can all be done
within the spin coherence time.
Let us estimate the typical parameters. A
large magnetic field gradient can be generated by moving the nanodiamond
close to a magnetic tip.
Here we take the gradient $G=10^{5}$~T/m and get the coupling $%
\lambda \simeq 2\pi \times 52$~kHz for a nanodiamond with the diameter $d=30$%
~nm in an optical trap with a trapping frequency $\omega _{m}=2\pi \times
0.5 $~MHz. The Fock states and their superpositions can then be generated
with a time scale $1/\lambda $ about a few $\mu $s, and the QND detection
rate $2|\chi |\sim 2\pi \times 25$~kHz with the detuning $||\Omega |-\omega
_{m}/2|\sim 5\lambda $. The NV electron spin dephasing time over $1.8$ ms
has been observed at room temperature \cite{Bala09}, which is long compared
with the Fock state preparation time $1/\lambda $ and the detection time $%
1/\left( 2|\chi |\right) $. 

\subsection{Sch\"odinger's cat states}
Creating Shr\"{o}dinger's cat states with massive objects is one of the
most challenging and attractive goals in macroscopic quantum mechanics.\cite{Bassi13,Chen13,Nimmrichter13}.
To generate spatial quantum superpositions and other
non-Gaussian states with an optical cavity, however, requires a very strong
quadratic coupling \cite{Romero11,Romero11a,Thompson08}. This is a very demanding
requirement. To enhance the quadratic coupling, Romero-Isart \emph{et al.}
\cite{Romero11} proposed to
prepare spatial quantum superpositions of nanoparticles with two
inter-connected high-finesse optical cavities:
one cavity for ground state cooling, and the other cavity for preparing the superposition
state with a squared position measurement when the nanoparticle falls
through it. The Sch\"odinger's cat state can also be generated by
ultraviolet (UV) laser if the ground state is reached \cite{Kaltenbaek12}. Then, we turn off the optical trap, and let the
wavefunction expand for a time $t_1$. A tightly focused UV laser pulse is shot through the center of the expanded
wavefunction, whose scale is in the order of hundreds nanometers. The Schr\"odinger's cat state
will generate conditional on no light being scattered. We can repeat the procedures
until Schr\"odinger's cat state is generated.

Here we discuss how to create the Sch\"odinger's cat state with a levitated nanodiamond with a NV center \cite{Yin13}. As we discussed in the previous subsection, the strong nonlinear coupling can be realized in
the nanocrystal diamond with building in NV centers. Therefore, in this system,
spatial quantum superpostion state, or Schr\"odinger's cat state, can be easily generated without measurement.
Without the
microwave driving, the spin-mechanical coupling Hamiltonian takes the form
\begin{equation}
H=\hbar \omega _{m}a_{m}^{\dagger }a_{m}+\hbar \lambda
S_{z}(a_{m}+a_{m}^{\dagger }).  \label{eq:Hlow}
\end{equation}%
The mechanical mode is initialized to the vacuum state $|0\rangle_m $ (or
a Fock state $|n_{m}\rangle_m $) in a strong trap with the trapping frequency $%
\omega _{m0}$ and the NV center spin is prepared in the state $|0\rangle $.
Although the ground state cooling is most effective in a strong trap, to
generate large spatial separation of the wave packets it is better to first
lower the trap frequency by tuning the laser intensity for the optical trap.
While it is possible to lower the trap frequency through an adiabatic sweep
to keep the phonon state unchanged, a more effective way is to use a
non-adiabatic state-preserving sweep \cite{Chen10}, which allows arbitrarily
short sweeping time.
We denote $|n_{m}\rangle _{m1}$ as the mechanical state in the lower
frequency $\omega _{m1}$.

\begin{figure}[t]
\centering
\includegraphics[width=8.5cm]{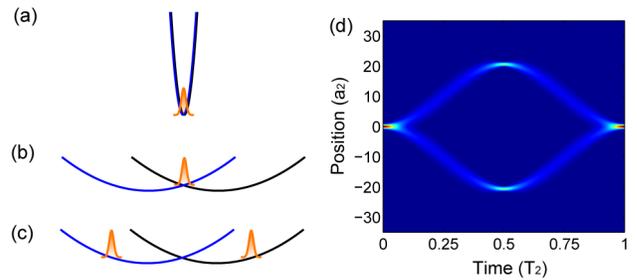}
\caption{ (a) Maximum spatial separation $D_{m}$ of the superposition state
as a function of trap frequency $\protect\omega _{m2}$ when the magnetic
gradient is $10^{5}$ T/m. (b) Maximum spatial separation $D_{m}$ as a
function of the magnetic gradient $G$ when the trapping frequency is 1 kHz.
Macroscopic superposition states with separation larger than the size of the
particle can be achieved with a moderate magnetic gradient. }
\label{fig:distance1}
\end{figure}

We then apply an impulsive microwave pulse to
suddenly change the NV spin to the state $(|+1\rangle +|-1\rangle )/\sqrt{2}$
and simultaneously decrease the trap frequency to $\omega _{m2}\leq \omega
_{m1}$. The evolution of the system state under the Hamiltonian (Eq. \ref{eq:Hlow}) then
automatically split the wave packet for the CoM motion of the
nanodiamond (see the illustration in Fig. \ref{fig:distance1}). The splitting attains the
maximum at time $T_{2}/2=\pi /\omega _{m2}$, where the maximum distance of
the two wave packets in the superposition state is $D_{m}=8\lambda
a_{2}/\omega _{m2}=4g_{s}\mu _{B}G/(m\omega _{m2}^{2})$, where $a_{2}=\sqrt{%
\hbar /2m\omega _{m2}}$. At this moment, the system state is $|\Psi
_{S}\rangle =(|+1\rangle |D_{m}/2\rangle _{n_{m}}+|-1\rangle
|-D_{m}/2\rangle _{n_{m}})/\sqrt{2}$, where $|\pm D_{m}/2\rangle
_{n_{m}}\equiv (-1)^{a_{m}^{\dagger }a_{m}}\exp \left[ \pm
D_{m}(a_{m}^{\dagger }-a_{m})/4a_{2}\right] \left\vert n_{m}\right\rangle
_{1}$ is the displaced Fock state (or coherent states when $n_{m}=0$).
This is just the entangled spatial superposition state.  
To transform the entangled cat state $|\Psi
_{S}\rangle $ to the standard cat state $\left\vert \psi _{\pm
}\right\rangle _{n_{m}}\equiv (|D_{m}/2\rangle _{n_{m}}\pm |-D_{m}/2\rangle
_{n_{m}})/\sqrt{2}$, we need to apply a disentangling operation to
conditionally flip the NV spin using displacement of the diamond as the
control qubit. This can be achieved as different displacements of the
wavepacket induce relative energy shifts of the spin levels due to the
applied magnetic field gradient. As an estimate, for the example that we consider
a $30$nm-diameter diamond in a $20$ kHz trap
under a magnetic gradient of $3\times 10^{4}$ T/m, the spin energy
splitting is about $2.4$ MHz between the $|+1\rangle |D_{m}/2\rangle _{n_{m}}
$ and $|-1\rangle |-D_{m}/2\rangle _{n_{m}}$ components, which is much
larger than the typical transition linewidth of the NV spin (in the order of
kHz). So we can apply first an impulsive microwave pulse to transfer the
component state $|+1\rangle |D_{m}/2\rangle _{n_{m}}$ to $|0\rangle
|D_{m}/2\rangle _{n_{m}}$ without affecting $|-1\rangle |-D_{m}/2\rangle
_{n_{m}}$ and then another pulse to transfer $|-1\rangle |-D_{m}/2\rangle
_{n_{m}}$ to $\pm |0\rangle |-D_{m}/2\rangle _{n_{m}}$. After the two
pulses, the spin state gets disentangled and the position of the diamond is
prepared in the quantum superposition state $\left\vert \psi _{\pm
}\right\rangle _{n_{m}}$.

To detect spatial superposition state, we can turn off the optical trap and
let the spatial wave function freely evolve for some time $t$. The split
wave packets will interference just like the Young's double slit experiment.
The period of the interference pattern is $\Delta z=2\pi \hbar t/(mD_{m})$.
As an estimation of typical parameters, we take $%
\omega _{m1}=\omega _{m2}=2\pi \times 20$ kHz, $d=30$ nm, and magnetic field
gradient $3\times 10^{4}$ T/m. The spin-phonon coupling rate $\lambda \simeq
2\pi \times 77$ kHz and the maximum distance $D_{m}\simeq 31 a_2$. The
preparing time of sperposition state is about $25$ $\mu $s, which is much
less than the coherence time of the NV spin. For the time of flight
measurement after turn-off of the trap, we see the interference pattern with
a period of $47$ nm after $t=10$~ms,  which is large
enough to be spatially resolved \cite{Li10,Li11,Gieseler12}.

\section{Ultrasensitive force detection}

In an ultra-high vacuum environment, the CoM motion of optically levitated sensors experiences minimal dissipation, enabling ultra-sensitive force detection \cite{Yin11,Geraci10,rugar,yocto}. Unlike conventional sensors consisting of solid-state mechanical resonators, e.g. cantilevers or membranes, the CoM motion of optically trapped dielectric objects is immune to the chief sources of dissipation in these devices at low pressure, consisting of lossy internal flexural and vibrational modes, surface imperfections, and clamping mechanisms.  The result is sub-attonewton force sensitivity that may have a number of applications ranging from Casimir force measurements, experimental gravitation, electric or magnetic field sensing, single molecules detecting, to inertial sensing.

\subsection{Force sensing with mechanical oscillators.}
High force sensitivity resonant sensors have typically consisted of solid-state micro-fabricated structures, for example cantilever beams or membranes \cite{rugar,teufel}. The achievement of aN/Hz$^{1/2}$ sensitivity in cryogenic cantilevers has lead to magnetic resonance force microscopy with the sensitivity to detect single electron spins in solids \cite{rugar2}, and has allowed sensitive tests for non-Newtonian gravity at the $\sim 10 \mu$m length scale \cite{stanford}.  In these systems, the internal materials losses and clamping mechanisms are responsible for limiting the quality factor of the oscillator to typically below $Q \sim 10^6$.   For force detection, it is desirable to have minimal dissipation, as the minimum detectable force due to thermal noise scales as $Q^{-1/2}$.  For a harmonic oscillator with natural frequency $\omega_0$ it can be expressed as
\begin{equation}
\label{fmin}
F_{\rm{min}}  = [4 k k_B T b/ \omega_0 Q ]^{1/2},
\end{equation}
where $b$ is the bandwidth of the measurement, $T$ is the effective temperature of the mode under consideration, and $k$ is the spring constant.  In ultra-high vacuum, the CoM motion of optically levitated micron-sized dielectric spheres and could exhibit $Q$ factors approaching $10^{12}$, leading to force sensitivity well below $1$ aN/Hz$^{1/2}$ at room temperature.  Experiments performed thus far have achieved inferred force sensitivities at the level of $\sim 10^{-20}$ N/Hz$^{1/2}$ for a $70$ nm particle at $P=10^{-5}$ mbar in Ref. \cite{Gieseler12}, and of order $10^{-19}$ N/Hz$^{1/2}$ for a $3$ $\mu$m diameter sphere feedback-cooled to 1.5 mK at 5.2 mPa in Ref. \cite{Li11}.

For a particle of mass $m$ in an optical trap, we can rewrite Eq. (\ref{fmin}) as $
F_{\rm{min}}  = [4 k_B T m b \gamma_g ]^{1/2},$
where the background gas
collision has a loss rate of $\gamma_g=16 P_{\rm{gas}}/(\pi \bar{v}
\rho a)$ \cite{epstein}, for a background air pressure of
$P_{\rm{gas}}$ and rms gas velocity $\bar{v}$, and a sphere of radius $a$ and density $\rho$.  However, as we discuss below, this formula must be modified due to heating by the recoil of scattered trap laser photons. Such scattering produces a heating rate $\gamma_{sc} = \frac{2}{5} \frac{\pi^2 \omega_0 V}{\lambda^3} \frac{(\epsilon-1)}{(\epsilon+2)}$, where $V$ is the sphere volume, $\lambda$ is the trap laser wavelength, and $\epsilon$ is the real part of the dielectric function for the sphere.

Laser cooling is essential for operation in high vacuum for several reasons. A mechanical oscillator with frequency $\sim 100$ kHz and $Q=10^{12}$ will respond to perturbations with a characteristic time
scale of $2Q/\omega_0 \approx 3 \times 10^6$ s, which is not practical for laboratory measurement. The cooling thus serves to damp the motion of the oscillator so that perturbations to the system can ring-down within reasonably short periods of time.  In addition, as the laser intensity determines the trapping frequency, it must be stabilized if the particle is to remain on resonance in the case of resonant detection. By also damping the oscillator, the laser cooling can therefore significantly reduce the requirement on the laser intensity stabilization.
Finally, the cooling is necessary to mitigate heating due to the recoil of trap laser photons. Such recoil heating leads to a momentum diffusion process, which left unchecked, can result in heating of the CoM motion of the particle and its eventual loss.  This heating modifies the expected form of the thermal noise limited force sensitivity.

Either (active) feedback cooling or (passive) cavity cooling serves to damp the $Q$ factor to $Q_{\rm{eff}}$ while at the same time the mode temperature is reduced to $T_{\rm{eff}}$.  The force sensitivity scales as $\sqrt{T_{\rm{eff}}/Q_{\rm{eff}}}$: the minimum detectable force due to thermal noise at temperature
$T_{\rm{eff}}$ is $F_{\rm{min}} = \sqrt{\frac{4 k k_B T_{\rm{eff}}
b}{\omega_0 Q_{\rm{eff}}}},$
where $k$ is the CoM mode spring
constant, and $b$ is the bandwidth of the measurement.

For example, in the case of cavity-cooling, the thermal-noise limited minimum detectable force becomes
\begin{equation}
F_{\rm{min}} = \sqrt{4 k_B T m b \gamma_g \left[1+\frac{\gamma_{\rm{sc}}+R_{+}}{n_i \gamma_g}\right]}.
\label{heq}
\end{equation}

We can define a factor $\chi = \frac{\gamma_{\rm{sc}}+R_{+}}{n_i \gamma_g}$ which describes the importance of photon recoil heating $\gamma_{\rm{sc}}$ and the efficiency of the cavity cooling. Here $n_i \equiv k_B T / \hbar \omega_0$ is the initial mean phonon occupation number. The factor $R_{+}$, defined in Ref. \cite{Chang10} can be minimized by going into the resolved sideband regime and can be generally neglected when compared with $\gamma_{sc}$.  There are two general regimes of scaling, $\chi << 1$ and $\chi >>1$.  For $\chi <<1$, the effects of photon recoil do not significantly degrade the force sensitivity, and $F_{\rm{min}} \propto a T^{1/4} P^{1/2}$ and is independent of trap frequency.  In the regime $\chi >>1$, photon recoil heating becomes significant, and damping without an equal amount of cooling occurs.  Here for a nanosphere the sensitivity scales as $F_{\rm{min}} \propto \omega_0 a^3 $ and is independent of $T$ and $P$.   A micro-disc geometry scatters much less light, as pointed out in Ref. \cite{kimblemem} and recoil heating is significantly reduced.  Also the micro-disc allows a larger mass to be trapped and localized in a particular anti-node of the standing wave in the cavity which is advantageous e.g. for gravitational wave strain sensitivity \cite{Arvanitaki13}.

\begin{figure}[!t]
\begin{center}
\includegraphics[width=0.9\columnwidth]{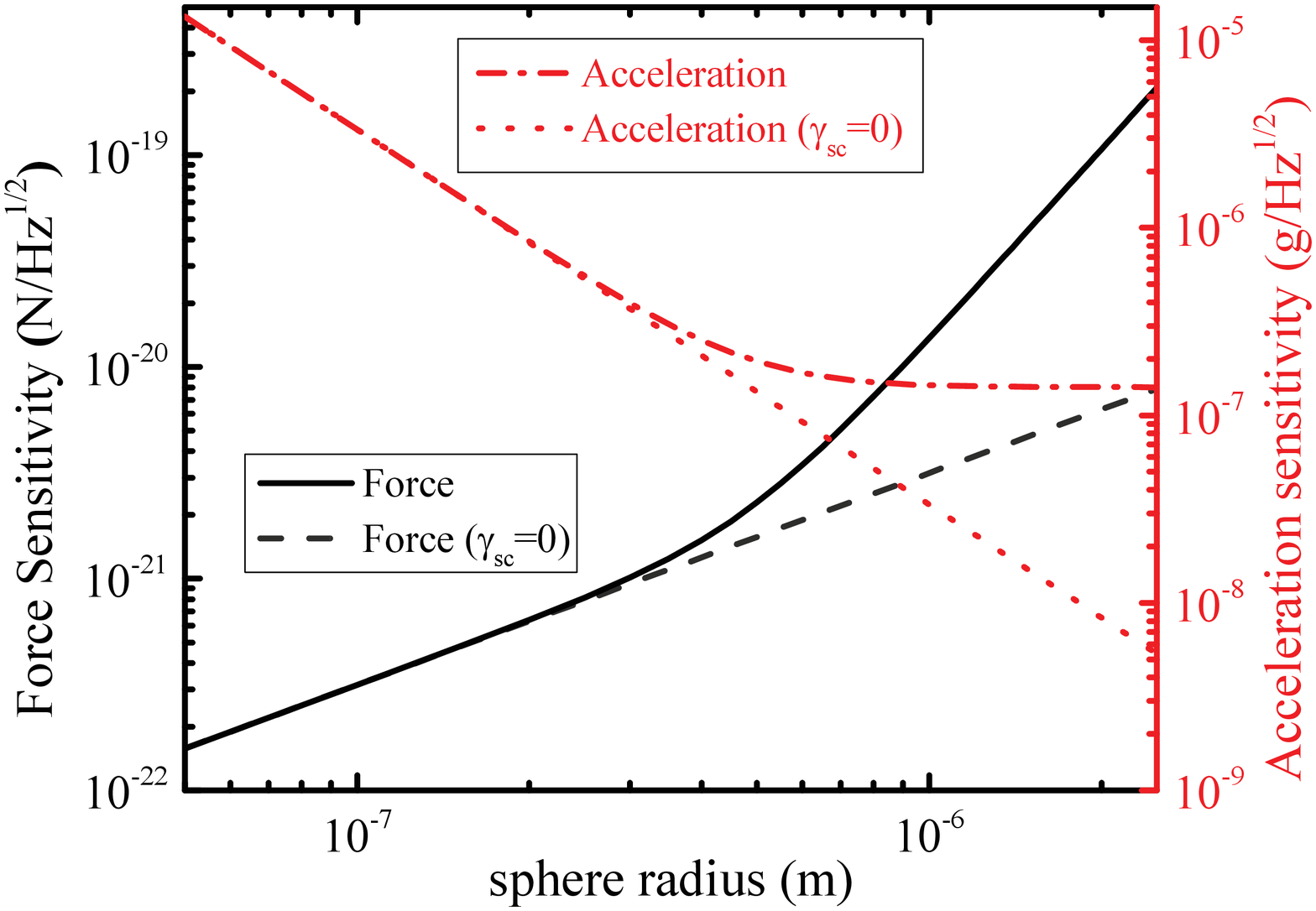}
\caption{Thermal noise limited force sensitivity and acceleration sensitivity for an optically trapped silica microsphere at pressure $P = 10^{-10}$ Torr and $T = 300$ K versus sphere radius at $1$ kHz trap frequency. The deviation from the scaling in Eq. (\ref{fmin}) is due to photon recoil heating.
\label{forceplot1}}
\end{center}
\end{figure}

\begin{figure}[!t]
\begin{center}
\includegraphics[width=0.9\columnwidth]{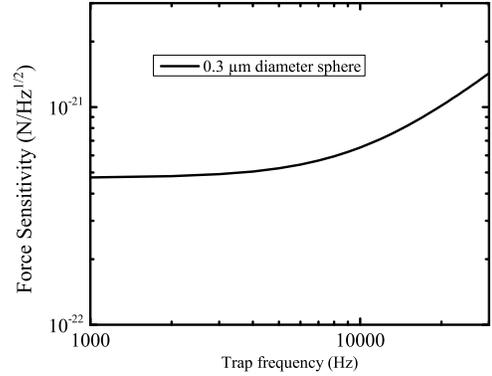}
\caption{Scattering-limited sensitivity for a $0.3$ $\mu$m diameter bead vs. trapping frequency for planned experimental parameters.
\label{forceplot2}}
\end{center}
\end{figure}

In Fig. \ref{forceplot1} we show the dependence of the force and acceleration sensitivity on the radius of the sphere at fixed trapping frequency of $1$ kHz assuming $R_{+} << \gamma_{sc}$, at $T = 300$ K and $P = 10^{-10}$ torr.  At larger radii the regime $\chi >>1$ is realized, with the minimum detectable force scaling as the sphere volume.  In Fig. \ref{forceplot2} we show the dependence of the force sensitivity on trapping frequency for the fixed size $a = 150$ nm. As the trapping frequency increases, the linear scaling with $\omega_0$ is apparent as the regime $\chi >>1$ is realized.

\subsection{Applications.} For the following discussion, we consider a dielectric sphere optically trapped and cooled in a cavity using two light fields of wavevector $k_t=2\pi/\lambda_{\rm{trap}}$ and $k_c=2\pi/\lambda_{\rm{cool}}$, respectively. The sphere is levitated in an anti-node of the trapping light which can be located near one of the mirrors of the cavity at frequencies ranging from $\sim 1-100$ kHz. The Gaussian profile of the trapping beam near the mode waist provides transverse confinement. Cooling of the transverse motion can be done with active feedback to modulate the power of a transverse cooling laser using the signal
from a transverse position measurement.

For detecting the position of the sensor, the phase of the cooling light reflected
from the cavity is modulated through the
optomechanical coupling $\partial{\omega_c}/\partial{z}=k_c g$.
Photon shot-noise limits the minimum detectable phase shift to
$\delta \phi \approx 1/(2\sqrt{I})$ where $I \equiv P_c /(\hbar
\omega_c)$ \cite{hadjar}. The corresponding photon shot-noise
limited displacement sensitivity
 is $ \sqrt{S_z(\omega)}= \frac{\kappa}{4k_cg}
\frac{1}{\sqrt{I}}\sqrt{1+\frac{4\omega^2}{\kappa^2}}$
\cite{kippenbergdisp}, for an impedance matched cavity. Here $P_c$ and $\omega_c$ are the cooling laser power and frequency, $g=\frac{3V}{4V_c} \frac{\epsilon-1}{\epsilon+2} \omega_c$, and $\kappa$ is the optical cavity loss rate. The cavity mode volume is $V_c$. The thermally-driven resonant motion of the sensor is typically much greater than the photon shot noise limited detection floor due to the high $Q_{\rm{eff}}$.

\subsubsection{Short-distance tests of gravity.} By trapping a nanosphere at an anti-node located at sub-micron distance from one of the cavity mirrors, it is possible to realize an experiment for testing gravity at the micron length scale \cite{Geraci10}.  Non-Newtonian gravity-like forces can be tested by monitoring the displacement of the sphere as a mass is brought behind the cavity mirror. Short-range corrections to Newtonian gravity are generally
parameterized according to a Yukawa-type potential
\begin{equation}
V=-\frac{G_Nm_1m_2}{r}\left[1+\alpha e^{-r/\lambda}\right],
\label{graveq}\end{equation}
where $m_1$ and $m_2$ are two masses interacting
at distance $r$, $\alpha$ is the strength of the correction relative
to gravity, and $\lambda$ is the range of the interaction. For two masses with density $\rho$ and linear dimesion $\lambda$ that are separated
by $r \approx \lambda$, a Yukawa-force scales roughly as
$F_Y \sim G_N \rho^2 \alpha \lambda^4$, rapidly decreasing with
smaller $\lambda$.  For a gold masss, for an
interaction potential with $\alpha=10^5$ and $\lambda=1$ $\mu$m,
$F_Y \sim 10^{-21}$ N. As the thermal-noise-limited
force sensitivity of micron-size trapped spheres can be of order $\sim
10^{-21}$ N$/\sqrt{{\rm{Hz}}}$, this setup therefore allows probing deep
into unexplored regimes.  For instance, current experimental limits
at $\lambda=1$ $\mu$m have only ruled out interactions with $|\alpha|$
exceeding $10^{10}$ \cite{Geraci10,adelberger,newlamoreaux,masuda}.

\subsubsection{Casimir Forces.} The Casimir Effect \cite{casimir} is a macroscopic manifestation of quantum vacuum fluctuations, and is a testament to the theory of quantum electrodynamics, arguably the most accurately known theory in physics.  At the same time, developing our understanding of it is becoming essential for pushing the size limits in nanotechnology and nano-electro-mechanical systems.   One of the most widely studied geometries involves the Casimir interaction between a sphere and plane. Previous measurements have been performed in the limit that their separation distance $d$ is small compared with the sphere radius $a$ \cite{sushkov,decca,lamoreaux,decca2,casreview}.  On the other hand, the Casimir-polder limit has also been explored using cold atoms, where the atomic size is much smaller than their distance to the plane \cite{hindscornell}.  However, there is a completely unexplored intermediate regime where the size of the sphere is on the order of the sphere-surface separation.  Such a regime poses an experimental challenge for commonly used measurement approaches involving a sphere attached to a torsional resonator – the mechanical resonator which is used for force sensing is tethered to the sphere, and therefore affects the geometry once the separation distance approaches the size of the sphere.
By using an optically-trapped nanosphere as the force sensor, one inherently overcomes this difficulty.  With a sphere trapped in an anti-node close to an end-mirror of the cavity, Casimir forces due to the metallic end-mirror can be measured as a frequency shift of the oscillator.  This type of experiment could allow a pristine dielectric-sphere/metal-plate geometry to be explored over a range of distances, from the short range limit where the proximity force approximation (PFA) is valid and the force varies as $1/d^3$, to the long range $1/d^5$ Casimir-polder limit \cite{Geraci10}.

\subsubsection{Gravitational Waves.} Nano- and micro-scale dielectric sensors trapped inside a medium-finesse optical cavity can be used to detect high frequency gravitational wave (GW) radiation \cite{Arvanitaki13}. The direct detection of gravitational radiation is very likely to occur in the next decade with the new generation of laser-interferometer gravitational wave observatories \cite{ligo,advligo,virgo,geo,LCGT}.  While these detectors have been optimized in the frequency band of $10-10^4$ Hz, their sensitivity decreases at higher frequency due to photon shot noise. The optically trapped sensor offers improved sensitivity in the frequency range of $50-300$ kHz using an approach that does not rely on a shot-noise limited displacement measurement of test mass mirrors, but rather depends on a precision force measurement on the resonant harmonically trapped sensor. The detector can yield sensitivities improved by more than an order of magnitude in this frequency band when compared with existing interferometers, while being only a fraction of their size.  The approach extends the effective search volume for sources between $100$ and $300$ kHz by $\sim 10 - 10^3$ when compared with Advanced LIGO \cite{advligo}. At such high frequencies, there may be sources of gravitational radiation from physics beyond the standard model. One example may result from the effects of the QCD axion on stellar mass black holes (BHs) through BH superradiance \cite{BH}. This novel signal comes from axion annihilation to gravitons and is monochromatic and long-lived.

In the approach proposed in Ref. \cite{Arvanitaki13}, a dielectic nanosphere or microdisc is optically trapped in an anti-node of a cavity of length at a position close to the input mirror. A second light field with two different frequency components is used to cool and read out the axial position of the levitated object, respectively. A passing gravitational wave displaces the sensor from its equilibrium position in the cavity, resulting in a measurable displacement of the levitated object.  Gravitational wave strain sensitivity can approach $ \sim 10^{-22}/\sqrt{\rm{Hz}}$ for frequencies near $100$ kHz for micron-sized discs in a cavity of length $100$ m. The resulting displacement of the sensor is resonantly enhanced when the frequency of the gravitational wave coincides with the optical trap frequency.

\subsubsection{Detecting single molecule collisions}
%

Detection of individual collisions between single
molecules and the nanosphere would lead to a test of the
Maxwell-Boltzmann distribution on single-collision level.
By using the 3D cooling sideband cooling scheme, we may
archive it by detecting the output light pulses.
Considering the gas pressure $P$ at temperature $T_\mathrm{env}$,
the radius of the sphere $r$, the molecule mass $m_m$, we have the
collision number per second  $N= (2\pi r^2)P/\sqrt{\pi m_m k_B
T_\mathrm{env}/2}$
, where $k_B$ is the
Boltzmann constant. The collision time is estimated to be much less
than the nanosphere oscillation time scale. The three phonon modes
initially in vacuum will be in a state with mean phonon number
$n_{j0}$: $\langle a^\dagger_j(t_0) a_j (t_0)\rangle= n_{j0}$ after
a single collision, where $t_0$ is the time when collision happens.
For this case, the output field is
$$a_{cj}^{\mathrm{out}}(t) = -i \frac{2g\alpha_j}{\sqrt{\kappa_j}} \exp
[-\frac{2g_j^2 |\alpha_j|^2}{\kappa_j}(t-t_0)]a_j(t_0) +
a_{cj}^{\mathrm{in}},$$
 It is easy to
find that $\int_{t_0}^\infty\langle a_{cj}^{\mathrm{out}}(t)
a_{cj}^{\mathrm{out}\dagger}(t)\rangle dt =n_{j0}$. This implies
that the output-pulse photon number is equal to the increase of the
phonon number after the collision. From above discussion, we get the
phonon decay time $\tau_j = \kappa_j/(4g^2_j |\alpha_j|^2)$, which
is also the pulse duration of the output light of mode $a_{cj}$. The
phonon number can be measured by detecting the output light pulse.
Therefore, $\tau_j$ is the measurement time for the phonon mode
$a_j$ after the collision. Therefore, as long as $\tau_j \ll 1/N$,
the collision events can be measured individually.

 Moreover, to make
detecting efficiency high, the phonon number
after the collision requires to be more than one.
For the first case, we suppose the collision is completely elastic. 
The average increase of the phonon number for $a_j$ is
$ n_{j0}=2m_m^2 \langle v_j^2\rangle/(\hbar \omega_j m)$ with
$\langle v^2_j\rangle$ the the mean velocity square along the axis
$q_j$. As a result, the requirement for the phonon number change
could be rewritten as $2 k_BT_{\mathrm{env}}> \hbar\omega_j
(m/m_m)$. If the collision is completely inelastic, the molecule
will attach on the surface of the nanosphere for a while before
being kicked out. The output velocity
distribution is completely determined by the temperature of the
nanosphere surface. The criteria should be either
$k_BT_{\mathrm{env}}> 2\hbar\omega_j (m/m_m)$, or
$k_BT_{\mathrm{sur}}> 2\hbar\omega_j (m/m_m)$, where
$T_{\mathrm{sur}}$ is the temperature of the surface of the
nanosphere. To distinguish elastic and inelastic collision, we can
cool the temperature to the limit that $k_BT_{\mathrm{env}}\ll
\hbar\omega_j (m/m_m)$, and makes the condition
$k_BT_{\mathrm{sur}}> 2\hbar\omega_j (m/m_m)$ fulfills by adding a
long wavelength laser to heat the sphere.  If the collisions are all
elastic, there is no signal on the photon detectors. If there are
parts of the collisions are inelastic, there are output pulses of
lights. Besides, the distribution of the photon numbers is
determined by the surface temperature of the sphere. In other words,
we can measure the surface temperature of the nanosphere by
detecting the output light pulses.

\subsubsection{Other applications.} By carrying a non-zero net electric charge, an optically trapped dielectric sphere becomes a sensitive detector for electric fields. For a charged sphere of diameter $300$ nm with an electric field of $\sim 10^7$ V/m at its surface, a $10^{-21}$ N/$\sqrt{\rm{Hz}}$ sensitivity corresponds to an electric field sensitivity of $\sim 10 \mu$V/m/$\sqrt{\rm{Hz}}$.  Correspondingly if the sphere were functionalized with a magnetic moment, sensitive magnetic field sensing may be possible, for example enabling magnetic resonance force microscopy \cite{rugar2}.


\begin{thebibliography}{99}
\bibitem{Penrose96}
R~Penrose.
\newblock On gravity's role in quantum state reduction.
\newblock {\em Gen. Rel. Grav.}, {\bf 28}(5), 1572 (1996).

\bibitem{christian2005}
J.~Christian.
\newblock Testing gravity-driven collapse of the wave function via cosmogenic
  neutrinos.
\newblock {\em Phys. Rev. Lett.} {\bf 95}, 160403 (2005).

\bibitem{wezel2008}
J.~{van} Wezel, T.~Oosterkamp, and J.~Zaanen.
\newblock Towards an experimental test of gravity-induced quantum state
  reduction.
\newblock {\em Phil. Mag.} {\bf 88}, 1005  (2008).

\bibitem{diosi1989}
L.~Di{\'{o}}si.
\newblock Models for universal reduction of macroscopic quantum fluctuations.
\newblock {\em Phys. Rev. A} {\bf 40}, 1165 (1989).

\bibitem{ghirardi1986}
G.~C. Ghirardi, A.~Rimini, and T.~Weber.
\newblock Unified dynamics for microscopic and macroscopic systems.
\newblock {\em Phys. Rev. D} {\bf 34}, 470 (1986).

\bibitem{Bassi13}
A.~{Bassi}, K.~{Lochan}, S.~{Satin}, T.~P. {Singh}, and H.~{Ulbricht}.
\newblock {Models of wave-function collapse, underlying theories, and
  experimental tests}.
\newblock {\em Reviews of Modern Physics} {\bf 85}, 471--527 (2013).

\bibitem{Nimmrichter13}
Stefan Nimmrichter and Klaus Hornberger.
\newblock Macroscopicity of mechanical quantum superposition states.
\newblock {\em Phys. Rev. Lett.} {\bf 110}, 160403 (2013).

\bibitem{Chen13}
Y.~{Chen}.
\newblock {Macroscopic quantum mechanics: theory and experimental concepts of
  optomechanics}.
\newblock {\em Journal of Physics B Atomic Molecular Physics} {\bf 46}, 104001
(2013).

\bibitem{Aspelmeyer2010}
M.~Aspelmeyer, S.~Groeblacher, K.~Hammerer, and N.~Kiesel.
\newblock Quantum optomechanics - throwing a glance.
\newblock J. Opt. Soc. Am. B {\bf 27}, A189-A197 (2010).

\bibitem{Wilson07}
I.~{Wilson-Rae}, N.~{Nooshi}, W.~{Zwerger}, and T.~J. {Kippenberg}.
\newblock {Theory of Ground State Cooling of a Mechanical Oscillator Using
  Dynamical Backaction}.
\newblock {\em Phys. Rev. Lett.}  {\bf 99}(9), 093901 (2007).

\bibitem{Marquardt07}
F.~{Marquardt}, J.~P. {Chen}, A.~A. {Clerk}, and S.~M. {Girvin}.
\newblock {Quantum Theory of Cavity-Assisted Sideband Cooling of Mechanical
  Motion}.
\newblock {\em Phys. Rev. Lett.} {\bf 99}(9), 093902 (2007).

\bibitem{Chan11}
J.~{Chan}, T.~P.~M. {Alegre}, A.~H. {Safavi-Naeini}, J.~T. {Hill}, A.~{Krause},
  S.~{Gr{\"o}blacher}, M.~{Aspelmeyer}, and O.~{Painter}.
\newblock {Laser cooling of a nanomechanical oscillator into its quantum ground
  state}.
\newblock {\em Nature}  {\bf 478}, 89--92 (2011).

\bibitem{Teufel11}
J.~D. {Teufel}, T.~{Donner}, D.~{Li}, J.~W. {Harlow}, M.~S. {Allman},
  K.~{Cicak}, A.~J. {Sirois}, J.~D. {Whittaker}, K.~W. {Lehnert}, and R.~W.
  {Simmonds}.
\newblock {Sideband cooling of micromechanical motion to the quantum ground
  state}.
\newblock {\em Nature}, {\bf 475}, 359--363 (2011).

\bibitem{Connell2010}
AD~O{\textquoteright}Connell and {\em et al.}
\newblock Quantum ground state and single-phonon control of a mechanical
  resonator.
\newblock {\em Nature} {\bf 464}, 697--703 (2010).

\bibitem{Aspelmeyer13}
M.~{Aspelmeyer}, T.~J. {Kippenberg}, and F.~{Marquardt}.
\newblock {Cavity Optomechanics}.
\newblock {\em ArXiv e-prints} 1303.4976, March 2013.

\bibitem{Chang10}
D.~E. Chang and {\em et al.}
\newblock {Cavity opto-mechanics using an optically levitated nanosphere}.
\newblock {\em PNAS} {\bf 107}(3), 1005--1010 (2010).

\bibitem{Romero10}
O.~Romero-Isart, ML~Juan, R.~Quidant, and JI~Cirac.
\newblock Towards quantum superposition of living organisms.
\newblock {\em New J. Phys.} {\bf 12}, 033015 (2010).

\bibitem{Li10}
Tongcang Li, Simon Kheifets, David Medellin, and Mark~G. Raizen.
\newblock {Measurement of the Instantaneous Velocity of a Brownian Particle}.
\newblock {\em Science}, {\bf 328}, 1673 (2010).

\bibitem{Li11}
T.~{Li}, S.~{Kheifets}, and M.~G. {Raizen}.
\newblock {Millikelvin cooling of an optically trapped microsphere in vacuum}.
\newblock {\em Nature Physics} {\bf 7}, 527--530 (2011).

\bibitem{Barker10}
P. F. Barker and M. N. Shneider.
Cavity cooling of an optically trapped nanoparticle.
{\em Phys. Rev. A} {\em 81}(2), 023826 (2010).

\bibitem{Barker10a}
P. F. Barker.
Doppler Cooling a Microsphere.
{\em Phys. Rev. Lett.} {\bf 105}(7), 073002 (2010).

\bibitem{Gieseler12}
J.~{Gieseler}, B.~{Deutsch}, R.~{Quidant}, and L.~{Novotny}.
\newblock {Subkelvin Parametric Feedback Cooling of a Laser-Trapped
  Nanoparticle}.
\newblock {\em Phys. Rev. Lett.} {\bf 109}(10), 103603 (2012).

\bibitem{Kiesel13}
N.~{Kiesel}, F.~{Blaser}, U.~{Delic}, D.~{Grass}, R.~{Kaltenbaek}, and
  M.~{Aspelmeyer}.
\newblock {Cavity cooling of an optically levitated nanoparticle}.
\newblock {\em arXiv e-prints} 1304.6679, 2013.

\bibitem{Asenbaum13}
P.~{Asenbaum}, S.~{Kuhn}, S.~{Nimmrichter}, U.~{Sezer}, M.~{Arndt}.
\newblock {Cavity cooling of free silicon nanoparticles in high-vacuum}.
\newblock {\em arXiv e-prints} 1306.4617, 2013.

\bibitem{Monteiro13}
T S Monteiro et al.
 Dynamics of levitated nanospheres: towards the strong coupling regime.
{\em New J. Phys.} {\bf 15}, 015001 (2013).

\bibitem{Yin09}
{Zhang-qi Yin}.
\newblock Phase noise and laser-cooling limits of optomechanical oscillators.
\newblock {\em Phys. Rev. A} {\bf 80}(3), 033821 (2009).

\bibitem{Yin11}
Z.-Q. {Yin}, T.~{Li}, and M.~{Feng}.
\newblock {Three-dimensional cooling and detection of a nanosphere with a
  single cavity}.
\newblock {\em Phys. Rev. A} 83(1), 013816 (2011).

\bibitem{Romero11}
O.~{Romero-Isart}, A.~C. {Pflanzer}, M.~L. {Juan}, R.~{Quidant}, N.~{Kiesel},
  M.~{Aspelmeyer}, and J.~I. {Cirac}.
\newblock {Optically levitating dielectrics in the quantum regime: Theory and
  protocols}.
\newblock {\em Phys. Rev. A} {\bf 83}(1), 013803 (2011).

\bibitem{Romero11a}
O.~{Romero-Isart}, A.~C. {Pflanzer}, F.~{Blaser}, R.~{Kaltenbaek}, N.~{Kiesel},
  M.~{Aspelmeyer}, and J.~I. {Cirac}.
\newblock {Large Quantum Superpositions and Interference of Massive
  Nanometer-Sized Objects}.
\newblock {\em Phys. Rev. Lett.} {\bf 107}(2), 020405 (2011).

\bibitem{Romero11b}
O.~{Romero-Isart}.
\newblock {Quantum superposition of massive objects and collapse models}.
\newblock {\em Phys. Rev. A} {\bf 84}(5), 052121 (2011).

\bibitem{Geraci10}
Andrew~A. Geraci, Scott~B. Papp, and John Kitching.
\newblock Short-range force detection using optically cooled levitated
  microspheres.
\newblock {\em Phys. Rev. Lett.} {\bf 105}(10), 101101 (2010).

\bibitem{PhysRevA.86.063809}
Wenjie Nie, Yueheng Lan, Yong Li, and Shiyao Zhu.
\newblock Effect of the casimir force on the entanglement between a levitated
  nanosphere and cavity modes.
\newblock {\em Phys. Rev. A}  {\bf 86}(6), 063809 (2012).

\bibitem{Arvanitaki13}
A.~{Arvanitaki} and A.~A. {Geraci}.
\newblock {Detecting High-Frequency Gravitational Waves with Optically
  Levitated Sensors}.
\newblock {\em Phys. Rev. Lett.} {\bf 110}(7), 071105  (2013).

\bibitem{Kaltenbaek12}
R.~{Kaltenbaek}, G.~{Hechenblaikner}, N.~{Kiesel}, O.~{Romero-Isart}, K.~C.
  {Schwab}, U.~{Johann}, and M.~{Aspelmeyer}.
\newblock {Macroscopic quantum resonators (MAQRO). Testing quantum and
  gravitational physics with massive mechanical resonators}.
\newblock {\em Experimental Astronomy} {\bf 34}, 123--164 (2012).

\bibitem{Arita11}
Y. Arita, A. W. McKinley, M. Mazilu, H. Rubinsztein-
Dunlop, and K. Dholakia.
\newblock Picoliter rheology of gaseous
media using a rotating optically trapped birefringent microparticle.
\newblock {\em Anal. Chem.} {\bf 83}(23), 8855 (2011).

\bibitem{Law12}
H. K. Cheung and C. K. Law, Optomechanical coupling between a moving dielectric sphere and radiation fields: A Lagrangian-Hamiltonian formalism.
{\em Phys. Rev. A} {\bf 86}, 033807 (2012).

\bibitem{Shi13}
H. Shi and M. Bhattacharyaa, Coupling a small torsional oscillator to large optical angular momentum.
{\em Journal of Modern Optics} {\bf  60}(5), 382-386 (2013); H. Shi and M. Bhattacharya, Mechanical memory
for photons with orbital angular momentum. {\em J. Phys. B: At. Mol. Opt. Phys.} {\bf 46}(15), 151001 (2013).

\bibitem{Lechner12} W. Lechner, S. J. M. Habraken, N. Kiesel, M. Aspelmeyer,
P. Zoller, {\em Phys. Rev. Lett.} \textbf{110}, 143604 (2013).

\bibitem{Habraken13} S.J.M. Habraken, W. Lechner, P. Zoller, {\em Phys. Rev. A}
\textbf{87}, 053808 (2013).

\bibitem{ashkin1971}
A.~Ashkin and J.~M. Dziedzic.
\newblock Optical levitation by radiation pressure.
\newblock {\em Appl. Phys. Lett.} {\bf 19}, 283 (1971).

\bibitem{Ashkin76}
A.~{Ashkin} and J.~M. {Dziedzic}.
\newblock {Optical levitation in high vacuum}.
\newblock {\em Appl. Phys. Lett.} {\bf 28}, 333 (1976).

\bibitem{Li13}
Tongcang Li, Fundamental tests of physics with optically trapped microspheres,
Springer, New York (2013).

\bibitem{Gieseler13}
J.~{Gieseler}, , L.~{Novotny}, and R.~{Quidant}.
\newblock {Thermal nonlinearities in a nanomechanical oscillator}.
\newblock {\em arXiv e-prints} 1307.4684, 2013.

\bibitem{einstein1907}
A.~Einstein.
\newblock Theoretische bemerkungen {\"{u}ber} die {Brownsche} bewegung.
\newblock {\em Zeit. f. Elektrochemie} {\bf 13}, 41 (1907).

\bibitem{cohadon1999}
P.~F. Cohadon, A.~Heidmann, and M.~Pinard.
\newblock Cooling of a mirror by radiation pressure.
\newblock {\em Phys. Rev. Lett.}, {\bf 83}(16), 3174-3177 (1999).

\bibitem{Isart12}
O. Romero-Isart, L. Clemente, C. Navau, A. Sanchez, and J. I. Cirac,
Quantum Magnetomechanics with Levitating Superconducting Microspheres.
{\em Phys. Rev. Lett.} {\bf 109}, 147205 (2012).

\bibitem{Cirio12}
M. Cirio, G. K. Brennen, and J. Twamley,
Quantum Magnetomechanics: Ultrahigh-Q-Levitated Mechanical Oscillators.
{\em Phys. Rev. Lett.} {\bf 109}, 147206 (2012).

\bibitem{PhysRevA.56.R1095}
T.~A. Savard, K.~M. O'Hara, and J.~E. Thomas.
\newblock Laser-noise-induced heating in far-off resonance optical traps.
\newblock {\em Phys. Rev. A} {\bf 56}(2), R1095--R1098 (1997).



\bibitem{PhysRevLett.99.160801}
Thomas Corbitt and {\em et al.}
\newblock Optical dilution and feedback cooling of a gram-scale oscillator to
  6.9 mk.
\newblock {\em Phys. Rev. Lett.} {\bf 99}(16), 160801 (2007).

\bibitem{PhysRevA.78.021801}
Lajos Di\'osi.
\newblock Laser linewidth hazard in optomechanical cooling.
\newblock {\em Phys. Rev. A} {\bf 78}(2), 021801 (2008).

\bibitem{PhysRevA.80.063819}
P.~Rabl, C.~Genes, K.~Hammerer, and M.~Aspelmeyer.
\newblock Phase-noise induced limitations on cooling and coherent evolution in
  optomechanical systems.
\newblock {\em Phys. Rev. A} {\bf 80}(6), 063819 (2009).

\bibitem{Pender12}
G.~A.~T. {Pender}, P.~F. {Barker}, F.~{Marquardt}, J.~{Millen}, and T.~S.
  {Monteiro}.
\newblock {Optomechanical cooling of levitated spheres with doubly resonant
  fields}.
\newblock {\em Phys. Rev. A} {\bf 85}(2), 021802 (2012).

\bibitem{YH09}
{Zhang-qi Yin} and {Y.-J. Han}.
\newblock Generating EPR beams in a cavity optomechanical system.
\newblock {\em Phys. Rev. A} {\bf 79}(2), 024301 (2009).

\bibitem{Yin13}
Z.-q. {Yin}, T.~{Li}, X.~{Zhang}, and L.~M. {Duan}.
\newblock {Raising Schr\"odinger's cat with a levitated
  nanodiamond through spin-opto-mechanical coupling}.
\newblock {\em ArXiv e-prints} 1305.1701, May 2013.

\bibitem{Horowitz12}
V.~R. {Horowitz}, B.~J. {Aleman}, D.~J. {Christle}, A.~N. {Cleland}, and D.~D.
  {Awschalom}.
\newblock {Electron spin resonance of nitrogen-vacancy centers in optically
  trapped nanodiamonds}.
\newblock {\em Proceedings of the National Academy of Science}
  {\bf 109}, 13493--13497 (2012).

\bibitem{Geiselmann13}
M.~{Geiselmann}, M.~L. {Juan}, J.~{Renger}, J.~M. {Say}, L.~J. {Brown},
  F.~J.~G. {de Abajo}, F.~{Koppens}, and R.~{Quidant}.
\newblock {Three-dimensional optical manipulation of a single electron spin}.
\newblock {\em Nature Nanotechnology} {\bf 8}, 175--179  (2013).

\bibitem{Neukirch13}
L.~P. {Neukirch}, J.~{Gieseler}, R.~{Quidant}, L.~{Novotny}, and A.~N.
  {Vamivakas}.
\newblock {Observation of nitrogen vacancy photoluminescence from an optically
  levitated nanodiamond}.
\newblock {\em arXiv e-prints} 1305.1515, May 2013.

\bibitem{Mamin07}
H.~J. {Mamin}, M.~{Poggio}, C.~L. {Degen}, and D.~{Rugar}.
\newblock {Nuclear magnetic resonance imaging with 90-nm resolution}.
\newblock {\em Nature Nanotechnology}  {\bf 2}, 301--306 (2007).

\bibitem{Rabl09}
P.~Rabl, P.~Cappellaro, M.~V.~Gurudev Dutt, L.~Jiang, J.~R. Maze, and M.~D.
  Lukin.
\newblock Strong magnetic coupling between an electronic spin qubit and a
  mechanical resonator.
\newblock {\em Phys. Rev. B} {\bf 79}(4), 041302  (2009).

\bibitem{Kolk12}
S.~{Kolkowitz}, A.~C. {Bleszynski Jayich}, Q.~P. {Unterreithmeier}, S.~D.
  {Bennett}, P.~{Rabl}, J.~G.~E. {Harris}, and M.~D. {Lukin}.
\newblock {Coherent Sensing of a Mechanical Resonator with a Single-Spin
  Qubit}.
\newblock {\em Science} {\bf 335}, 1603--1606  (2012).

\bibitem{Robledo11}
L.~{Robledo}, L.~{Childress}, H.~{Bernien}, B.~{Hensen}, P.~F.~A. {Alkemade},
  and R.~{Hanson}.
\newblock {High-fidelity projective read-out of a solid-state spin quantum
  register}.
\newblock {\em Nature}  {\bf 477}, 574--578 (2011).

\bibitem{Meekhof96}
D.~M. Meekhof, C.~Monroe, B.~E. King, W.~M. Itano, and D.~J. Wineland.
\newblock Generation of nonclassical motional states of a trapped atom.
\newblock {\em Phys. Rev. Lett.}  {\bf 76}(11), 1796--1799 (1996).

\bibitem{Bala09}
G.~{Balasubramanian}, P.~{Neumann}, D.~{Twitchen}, M.~{Markham}, R.~{Kolesov},
  N.~{Mizuochi}, J.~{Isoya}, J.~{Achard}, J.~{Beck}, J.~{Tissler},
  V.~{Jacques}, P.~R. {Hemmer}, F.~{Jelezko}, and J.~{Wrachtrup}.
\newblock {Ultralong spin coherence time in isotopically engineered diamond}.
\newblock {\em Nature Materials}  {\bf 8}, 383--387 (2009).

\bibitem{Thompson08}
J.~D. {Thompson}, B.~M. {Zwickl}, A.~M. {Jayich}, F.~{Marquardt}, S.~M.
  {Girvin}, and J.~G.~E. {Harris}.
\newblock {Strong dispersive coupling of a high-finesse cavity to a
  micromechanical membrane}.
\newblock {\em Nature}  {\bf 452}, 72--75  (2008).

\bibitem{Chen10}
Xi~Chen, A.~Ruschhaupt, S.~Schmidt, A.~del Campo, D.~Gu\'ery-Odelin, and J.~G.
  Muga.
\newblock Fast optimal frictionless atom cooling in harmonic traps: Shortcut to
  adiabaticity.
\newblock {\em Phys. Rev. Lett.} {\bf 104}(6), 063002 (2010).

\bibitem{rugar}
H.J. Mamin and D.~Rugar.
\newblock Sub-attonewton force detection at millikelvin temperatures.
\newblock {\em Appl. Phys. Lett.}  {\bf 79}, 3358 (2001).

\bibitem{yocto}
R. Maiwald {\it{et. al.}}, \newblock Stylus ion trap for enhanced access and sensing.
 \newblock {\em Nature Physics} {\bf{5}}, 551-554 (2009);
 M. Biercuk {\it{et. al.}}, \newblock Ultrasensitive detection of force and displacement using trapped ions.
 \newblock {\em Nature Nanotechnology} 5, 646-50 (2010).


\bibitem{teufel} J.D.Teufel, T. Donner, M.A. Castellanos-Beltran, J.W. Harlow, K.W. Lehnert,
Nanomechanical motion measured with an imprecision below that at the standard quantum limit.
{\em Nature Nanotech.} {\bf 4}, 820-823 (2009).

\bibitem{rugar2} D. Rugar {\it{et. al.}}, Single spin detection by magnetic resonance force microscopy.
{\em Nature} {\bf{430}}, 329-332 (2004).

\bibitem{stanford}
A. A. Geraci {\it{et. al.}}, Improved constraints on non-Newtonian forces at 10 microns. {\em Phys. Rev. D} {\bf{78}}(2), 022002 (2008).

\bibitem{epstein} P. S. Epstein, On the Resistance Experienced by Spheres in their Motion through Gases.
{\em Phys. Rev.} {\bf{23}}(6), 710-733 (1924).

\bibitem{kimblemem} D.E. Chang et. al., Ultrahigh-Q mechanical oscillators through optical trapping.
{\em New. J. Phys.} {\bf{14}}, 045002 (2012).


\bibitem{hadjar} Y. Hadjar {\it{et. al.}}, High-sensitivity optical measurement of mechanical Brownian motion.
{\em Europhys. Lett.} {\bf{47}}(5), 545-551 (1999).
\bibitem{kippenbergdisp} G. Anetsberger {\it{et. al.}}, Near-field cavity optomechanics with nanomechanical oscillators.
{\em Nature Physics} {\bf{5}}, 909-914 (2009).
\bibitem{adelberger} E. G. Adelberger, J. H. Gundlach, B. R. Heckel, S. Hoedl, and S. Schlamminger, Torsion balance experiments: A low-energy frontier of particle physics. {\em Prog. Part. Nucl. Phys} 62, 102-134 (2009).
\bibitem{newlamoreaux} A. O. Sushkov, W. J. Kim, D. A. R. Dalvit, and S. K. Lamoreaux, New Experimental Limits on Non-Newtonian Forces in the Micrometer Range.
 {\em Phys. Rev. Lett.} {\bf 107}(17), 171101 (2011).
\bibitem{masuda} M. Masuda and M. Sasaki, Limits on Nonstandard Forces in the Submicrometer Range.
{\em Phys. Rev. Lett.} {\bf{102}}(17), 171101 (2009).
\bibitem{casimir} H. B. G. Casimir, On the attraction between two perfectly conducting plates. {\em Proc. K. Ned. Akad. Wet.} {\bf 51},793 (1948).
\bibitem{sushkov} A.O. Sushkov et. al., Observation of the thermal Casimir force. {\em Nature Physics} {\bf 7}, 230-233 (2011).
\bibitem{decca} R. S. Decca et. al., Tests of new physics from precise measurements of the Casimir pressure between two gold-coated plates.
{\em Phys. Rev. D}  {\bf 75}, 077101 (2007).
\bibitem{lamoreaux} S. K. Lamoreaux, Demonstration of the Casimir Force in the 0.6 to 6$\mu$m Range.
{\em Phys. Rev. Lett.} {\bf 78}, 5 (1997).
\bibitem{decca2} R. S. Decca et. al., Constraining New Forces in the Casimir Regime Using the Isoelectronic Technique.
{\em Phys. Rev. Lett.} {\bf 94}, 240401 (2005).
\bibitem{casreview} A. Lambrecht and S. Reynuad, {\em arXiv eprint} 1112.1301v1 (2011).
\bibitem{hindscornell} C.I.Sukenik et.al., Measurement of the Casimir-Polder force.
{\em Phys. Rev. Lett.} {\bf 70}, 560-563 (1993); D. M. Harber et. al., Measurement of the Casimir-Polder force through center-of-mass oscillations of a Bose-Einstein condensate. {\em Phys. Rev. A} {\bf 72}, 033610 (2005).

\bibitem{ligo}
B.Abbott, {\it{et. al.}}, LIGO: the Laser Interferometer Gravitational-Wave Observatory. {\em Rep. Prog. Phys.} {\bf{72}}, 076901 (2009).
\bibitem{advligo}
G. M. Harry (for the LIGO Scientific Collaboration), Advanced LIGO: the next generation of gravitational wave detectors.
 {\em Class. Quantum Grav.} {\bf 27}, 084006 (2010).
\bibitem{virgo}
T. Accadia {\it{et. al.}},  Virgo: a laser interferometer to detect gravitational waves. {\em Journal of Inst.} {\bf 7}, P030012 (2012);
 The Virgo Collaboration, note VIR-027A-09 (2009).
\bibitem{geo}
H. Grote (for the LIGO Scientific Collaboration), The GEO 600 status. {\em Class. Quantum Grav.} {\bf 27} 084003 (2010); B. Willke {\it{et. al.}}, The GEO-HF project. {\it{ibid.}} {\bf 23}, S207 (2006).
\bibitem{LCGT}
K. Kuroda (for the LCGT Collaboration), Status of LCGT. {\em Class. Quantum Grav.} {\bf 27}, 084004 (2010).
\bibitem{BH}
  A.~Arvanitaki, S.~Dimopoulos, S.~Dubovsky, {\it{et. al.}}, String axiverse.
    {\em Phys.\ Rev.\ D} {\bf 81}, 123530 (2010);
  A. Arvanitaki and S. Dubovsky, Exploring the string axiverse with precision black hole physics.
{\em Phys. Rev. D} \textbf{83}, 044026 (2011).
\end{thebibliography}
 \end{document}